\documentclass[conference]{IEEEtran}
\IEEEoverridecommandlockouts
\usepackage{cite}
\usepackage{multirow}
\usepackage{amsmath,amssymb,amsfonts}
\usepackage[normalem]{ulem}
\usepackage{algorithm}[]
\usepackage[noend]{algorithmic}
\usepackage{graphicx}
\usepackage{textcomp}
\usepackage{xcolor}
\usepackage{verbatim}
\usepackage{subfigure}

\newtheorem{myDef1}{Definition}

\newtheorem{definition}{Definition}

\def\BibTeX{{\rm B\kern-.05em{\sc i\kern-.025em b}\kern-.08em
    T\kern-.1667em\lower.7ex\hbox{E}\kern-.125emX}}
\begin{document}

\title{Learn to Explore: on Bootstrapping Interactive Data Exploration with Meta-learning}
\makeatletter
\newcommand{\removelatexerror}{\let\@latex@error\@gobble}

\newcommand{\phase}[1]{%
  \vspace{-1.25ex}
  \Statex\leavevmode\llap{\rule{\dimexpr\labelwidth+\labelsep}{\phaserulewidth}}\rule{\linewidth}{\phaserulewidth}
  \Statex\strut\refstepcounter{phase}\textit{Phase~\thephase~--~#1}
  \vspace{-1.25ex}\Statex\leavevmode\llap{\rule{\dimexpr\labelwidth+\labelsep}{\phaserulewidth}}\rule{\linewidth}{\phaserulewidth}}

\makeatother

\author{Yukun Cao, Xike Xie, and Kexin Huang\\
University of Science and Technology of China\\
\{ykcho, huang\_1773\}@mail.ustc.edu.cn, xkxie@ustc.edu.cn
}

\maketitle

\begin{abstract}
  Interactive data exploration (IDE) is an effective way of comprehending big data, whose volume and complexity are beyond human abilities.
  The main goal of IDE is to discover user interest regions from a database through multi-rounds of user labelling.
  Existing IDEs adopt active-learning framework, where users iteratively discriminate or label the interestingness of selected tuples.
  The process of data exploration can be viewed as the process of training of a classifier, which determines whether a database tuple is interesting to a user.
  An efficient exploration thus takes very few iterations of user labelling to reach the data region of interest.
  In this work, we consider the data exploration as the process of few-shot learning, where the classifier is learned with only a few training examples, or exploration iterations. To this end, we propose a learning-to-explore framework, based on meta-learning, which learns how to learn a classifier with automatically generated meta-tasks, so that the exploration process can be much shortened.
  Extensive experiments on real datasets show that our proposal
outperforms existing explore-by-example solutions in terms of accuracy
and efficiency.

\end{abstract} 


\section{Introduction}

Interactive data exploration (IDE {\it in short}) \cite{10.1145/2723372.2731084} is at the frontline of big data management, which tackles data comprehensibility challenges caused by fast data accumulation and limited human ability. The problem of IDE is challenging, because: 1) user interest is intangible so that incremental refinement/exploration of user interests is required~\cite{10.1145/2588555.2610523}; 2) user interest is indescribable in the sense that it is often too complex to be specified by a user through traditional query languages (e.g., SQL)~\cite{10.1145/3318464.3383126}.

For example, Alice and Bob explore sky objects in the Sloan Digital Sky Survey (SDSS) database\footnote{https://www.sdss.org/}. Alice is an amateur astronomer,
and her familiar attributes are relatively limited, $\{rowc$, $colc$, $ra$, $dec\}$. However, her
data interest is so uncertain that it is hard for her to express accurately. Alternatively, she can browse and selectively label some database tuples so that the recommendation of tuples or queries reflecting her interests can be enabled by IDEs. Bob is an astronomical scientist whose data interest covers a wide range of attributes, $\{rowc$, $colc$, $ra$, $dec$, $sky\_u$, $sky\_g$, $...\}$\footnote{These attributes are the photometric attributes of sky objects. Details are in: https://skyserver.sdss.org/}. However, his requirements (e.g., mathematical expressions involving multiple attributes) are too complex to be expressed by conventional database queries, and even database experts take much time to write dedicated filters. But it is easy for Bob to label whether a specific data tuple meets his needs.

Following explore-by-example paradigm~\cite{10.1145/2588555.2610523,7539596,10.14778/3275536.3275542}, the main goal of IDE is to discover user interest regions (UIR {\it in short}) from a to-be-explored database through multiple rounds of user labelling.
The exploration process can be viewed as the training process of classifiers~\cite{10.1145/3318464.3383126}, deciding if a database tuple is ``interesting'' to a user.
The output of IDE refers to arbitrarily attainable data query regions, covering user interested tuples in the explored database.
Technically, the indescribability brings in the challenge of generality in UIR representation~\cite{10.1145/3318464.3383126}.
The intangibility brings in the challenge of ``slow convergence''. For example, hundreds of iterations of labelling is needed to converge to a UIR~\cite{10.1145/2588555.2610523}.

\begin{table}[th]
\vspace{-10pt}
\centering
\caption{Evolution of IDE under ``Explore-by-Example '' Paradigm}
\label{tab:ide}
\scriptsize
\begin{tabular}{cccc}
\hline
 & \textbf{UIR in subspace} & \textbf{Classifier} & \textbf{Techniques} \\ \hline
\textbf{AIDE~\cite{10.1145/2588555.2610523,7539596} } & Linear & Decision Tree & Active-Learning \\ \hline
\textbf{DSM~\cite{10.14778/3275536.3275542} } & Convex & SVM & Active-Learning \\ \hline
\textbf{LTE} & Arbitrary & Neural Networks & Meta-Learning \\ \hline
\end{tabular}
\end{table}

In general, a high-performance IDE is expected to achieve both high efficiency and accuracy. The efficiency refers to human efforts expended on the ``interestingness'' labelling.
The accuracy refers to the closeness between the inferred UIR and the real one. The pursuit of efficiency and accuracy can also be observed from IDE technology evolution, in Table~\ref{tab:ide}.
A better classifier leads to a faster convergence; and a better UIR representation leads to a higher accuracy, which in turn prompts the classifier training, i.e., exploration  efficiency.

It is thus a natural evolution of IDE classifiers, emanating from machine learning, e.g., decision trees~\cite{10.1145/2588555.2610523,7539596}, support vector machines (SVM {\it in short})~\cite{10.14778/3275536.3275542}, and taking shape in deep learning, e.g., neural networks (NN {\it in short}). The NN has good capabilities in capturing abstract feature representation in the manner of stacked layers, with a good match to user interest exploration which is intangible and indescribable.
Despite the potential accuracy, there are several challenges: the NN classifier relies on a large number of user labels for training, which implies more exploration iterations and human efforts, and thus slower convergence, contradicting the efficiency target. 

To this end, we propose to boost NN classifiers by meta-learning\footnote{A typical method such as MAML (Model-Agnostic Meta-Learning~\cite{10.5555/3305381.3305498} has the objective of learning an appropriate model initialization parameter in a range of meta-tasks.} for IDE systems.
The mechanism of meta-learning is also characterized as few-shot learning by literatures~\cite{hospedales2020metalearning}, which shows a good match to the cold start of data exploration, where labels are rare and precious.
We call the meta-learning supported NN classifier the {\it meta-learner}, which is pre-trained with automatically generated synthetic {\it meta-tasks}.
The pre-training equips meta-learners with good initialization parameters, so that they can be quickly adapted and generalized during the initial exploration phase.
Unlike conventional training strategies (e.g., active learning \cite{7539596, 10.14778/3275536.3275542}), it takes merely a few gradient optimization steps for classifiers to converge, corresponding to fewer iterations of user labelling, and therefore higher efficiency during the online exploration. From this point, the meta-learning process can be viewed as ``\underline{l}earning \underline{t}o \underline{e}xplore'' (LTE {\it in short}), which takes few-shot of labelling for a quality initialization of IDE.
The LTE framework has two features, in addition to high accuracy and efficiency.

\begin{figure}
    \centering \includegraphics[width=0.5\textwidth]{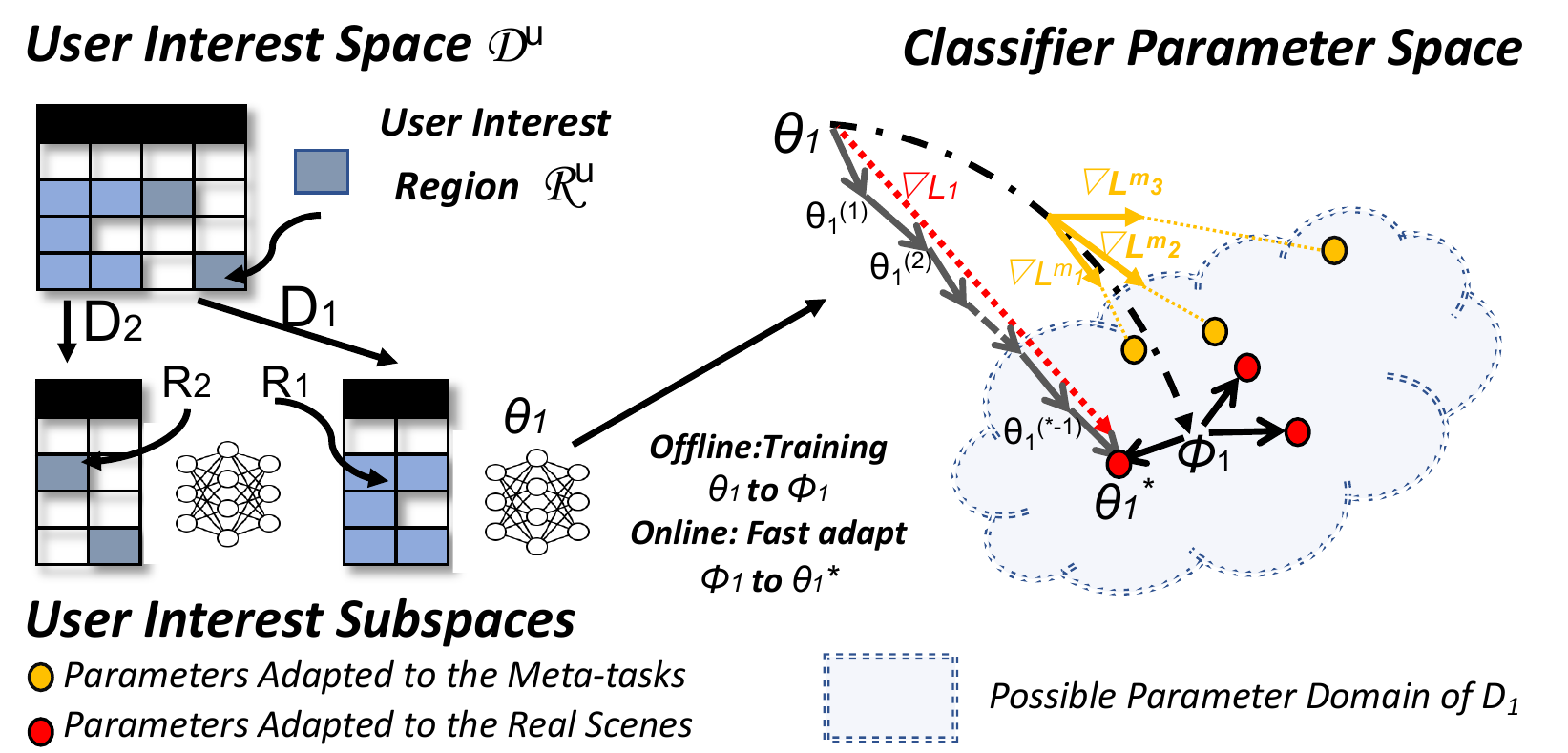}
    \vspace{-5pt}
    \caption{An Example of LTE}
    \label{frame_intro}
    \vspace{-5pt}
\end{figure}

First, the pre-training processing of meta-learners in LTE is unsupervised~\cite{hsu2018unsupervised}, so that it does not incur the overhead of user labelling.
Meanwhile, it extracts data distributions and insights from  lightweight sampled tuples set, represented as meta-knowledge, for initializing meta-learners with good parameters. Figure~\ref{frame_intro} shows the idea of meta-training.
Under regular training settings, a classifier's parameter $\theta_1$ would be trained by backpropagation which computes the gradient decent of a loss function, e.g., $\nabla L_1$, and goes through a long optimization path $\langle \theta_1^{(1)}, \theta_1^{(2)}, ..., \theta_1^{(*-1)} \rangle$ to converge to an optimal $\theta_1^*$.
It may not be cost efficient, since each gradient optimization step requires user labelling.
With meta-learning, the classifier parameter $\theta_1$ is pre-trained to $\phi_1$, which can be quickly adapted to parameter $\theta_1^*$ with much less optimization steps so as to reduce user labels, during the online exploration.

More, it supports concave or even scattered user interest regions. The motivation is towards tackling the indescribability challenge, under which user interests are somehow hard to be explicitly expressed in SQL templates or filters.
Figure~\ref{frame_intro} shows an example of UIR, which is shaded in a tabular dataset.
Given a multi-dimensional dataset, an IDE system ~\cite{10.14778/3275536.3275542} usually decomposes it into multiple low-dimensional datasets.
For example, data space $\mathcal{D}^u$ is decomposed into subspaces $D_1$ and $D_2$.
The projection of UIR on each subspace can be of arbitrary shapes. For example, $R_1$ is concave on $D_1$, and $R_2$ is a scattered region on $D_2$. Therefore, a powerful classifier is needed to deal with the generality setting of UIR.
Unlike existing works (in Table~\ref{tab:ide}), we do not make assumptions on UIR shapes.



Our main contributions are summarized as follows.
\begin{itemize}
  \item We propose, to our best knowledge, the first ``learn-to-explore'' framework, that harnesses meta-learning based neural network classifiers for data exploration.
  \item The meta-learner of the LTE framework is pre-trained with automatically generated meta-tasks, so that only a few gradient optimization steps are needed during the online exploration, leading to less exploration iterations and human efforts.
  \item Experiments on real datasets demonstrate that our proposal outperforms existing explore-by-example solutions in terms of accuracy
and efficiency. 
\end{itemize}

The rest of the paper is organized as follows.
Section~\ref{sec:relate} summarizes related works.
Section~\ref{sec:overview} presents basic concepts and the LTE framework.
Section~\ref{sec:meta_over} investigates the meta-learning process.
Section~\ref{sec:taskgen} studies the generation of meta-tasks.
Section~\ref{sec:meta-train} investigates the training process of meta-learners.
Section~\ref{sec:pre_opt} introduces other critical techniques of the LTE framework, i.e., tabular data preprocessing and few-shot optimization.
Section~\ref{sec:ret} reports experimental results and Section~\ref{sec:con} concludes the paper.

\section{Related Work}
\label{sec:relate}

\textit{\textbf{Interactive Data Exploration.}} Interactive Data exploration is about how users can extract knowledge from data using system assistance and interactive guidance, when they do not have exact query requirements \cite{10.1145/2723372.2731084,10.1145/3318464.3383126,new_trend_exp}.
Early works focus on simple user interactions. For example, \cite{4812548,6816674} requires users to gradually provide interest attribute values to drill down and finally return interest tuple set. Some works support the exploration with error guarantees and response deadlines for specific data types and query templates \cite{10.14778/2367502.2367533,10.14778/3402755.3402799,10.1145/2588555.2593666}. 
Some works study preparatory data exploration with the support of online analytical processing \cite{WasayWDI17,XieZHPJY20,GaoXZP22,XieHPJC16,BFb0100984}. In addition, there are works on the exploration result visualization \cite{visual,10.14778/2732240.2732250} and query formulation \cite{10.1145/2396761.2398507,10.14778/2367502.2367542}.

 Numerous recent researches aim to expand the diversity of exploration modes, especially by utilizing machine/deep learning to optimize/model various exploration modules. The ``explore-by-example'' systems\cite{7539596,10.14778/3275536.3275542} are designed to discover user interest regions through tuple-level user labeling and employ active learning to improve interaction efficiency. ``Insights" driven systems \cite{Extracting_Insights,QuickInsights,MetaInsight} formalize interesting patterns (including correlations, anomalies, trends, etc.) in multidimensional data as ``insights'' and propose some interactive exploration frameworks for ``insights''. Automated exploratory data analysis (EDA) systems recommend an exploration path for users, which generally requires predefined exploration modes and various types of user interactions.  Two representatives, ATENA\cite{Auto_EDA} and Dora\cite{dora}, utilize deep reinforcement learning to model the EDA process. ExplainED\cite{ExplainED} automatically generates semantic explanations for each step of the EDA process to guide the user's exploration, by natural language processing (NLP) techniques.
Our work is under the explore-by-example paradigm, which is considered as complementary systems to EDA systems \cite{Auto_EDA}.

In addition, some works are oriented toward specific exploration data types, such as graphs\cite{DiaoGMM21,HDAG-Explorer,MC-Explorer}, spatio-temporal data\cite{DeepTrack,Tabula,Efficient_map}, and time series \cite{Interactive_time,Visual_Exploration}. Some works \cite{Effortless,Steerable} focus on optimizing the visualization experience during interactive exploration.



\textit{\textbf{Explore-by-Example.}} IDEs under this paradigm~\cite{10.1145/2588555.2610523} originate from the research of ``query by example''~\cite{10.14778/2732269.2732273}, which recommends selective tuples in the databases as proxies for exploration targets. The latest IDE frameworks~\cite{7539596,10.14778/3275536.3275542} regard the exploration process as an incremental classification problem, and employ active learning to select the tuples that are most difficult to ``discriminate'' for users to label. However, due to the limitation of classifiers and the bottleneck of active learning, these frameworks focus on specific exploration targets.
For example, the state-of-the-art, DSM~\cite{10.14778/3275536.3275542}, assumes subspatial convexity and conjunctivity of UIRs.
Our work bootstraps the explore-by-example IDE paradigm, aided by meta-learning, for better exploration efficiency.


\textit{\textbf{Meta-Learning.}} It is often known as ``learning to learn'', which seeks to gain meta-knowledge from a set of machine learning tasks in order to improve the learning process~\cite{hospedales2020metalearning}. It belongs to the scene of few-shot learning~\cite{10.1145/3386252}.
We focus on a type of meta-learning method, which learns good initial parameters for meta-learners with meta-tasks.
 A typical meta-task (e.g., MAML~\cite{10.5555/3305381.3305498}) has both training and validation data, called \emph{support set} and {\it query set}, respectively. During the meta-training, the meta-learner iterates over the meta-tasks. At each iteration, a local learner is trained on the support set and tested by the query set. The meta-learner's parameters are then globally updated according to aggregated backpropagated loss measured by local testing errors.
Our work belongs to a challenging topic of unsupervised learning via meta-learning~\cite{hsu2018unsupervised}, since  meta-tasks are generated without label sets.

\section{Overview}
\label{sec:overview}

We introduce basic concepts in Section~\ref{subsec:concept},  show the framework overview in Section~\ref{subsec:framework}.


\subsection{Basic concepts}
\label{subsec:concept}

\textit{\textbf{User Interest Space.}} Suppose a database consisting of a set of attributes $A=\{a_1, ..., a_{|A|}\}$ and a set of $|A|$-dimensional tuples.
We define the domain space formed by attribute set $A$ as $\mathcal{D}=\{domain(a_{1})\times domain(a_{2}) \times ... domain(a_{|A|})\}$,
which covers all the database tuples.
A user $u$ is interested in a subset of attributes $A^u\subseteq A$, of which the domain space can be represented by $\mathcal{D}^u = \{domain(a^u_{1}) \times ... domain(a^u_{|A^u|})\}_{a^u_j \in A^u}$, called {\it user interested space}.

\textit{\textbf{User Interest Subspace.}}
The exploration target is to browse $\mathcal{D}^u$ for retrieving tuples interesting to user $u$.
Existing IDEs~\cite{10.14778/3275536.3275542,palma2020efficient} decompose
$\mathcal{D}^u$ into a set of disjoint low-dimensional {\it subspaces} $\{D_i\}_{i \leq n}$, where $\mathcal{D}^u = D_1 \times ...\times D_{n}$.


\textit{\textbf{User Interest Subregion (UIS).}}
Given a user interest subspace $D_i$, the UIS $R_i \subseteq D_i$ can be defined as the tuples $\{\tau\in D_i\}$ satisfying a user's interest. If user $u$'s exploration interest on $D_i$ is represented by a binary classifier $\mathcal{I}_i:D_{i} \rightarrow \{0,1\}$\footnote{Let $1$ be ``interesting'', and $0$ be ``not interesting''.}, UIS $R_i$ can be represented by
$R_i=\{\tau \in D_i|\mathcal{I}_i(\tau) =1\}$.

\textit{\textbf{User Interest Region (UIR).}}
Essentially, a user interest region $\mathcal{R}^u$ for user $u$ is the conjunctive combination of its subregions, $\mathcal{R}^u = \vee_{i \leq n}R_i$. The target of data exploration is to efficiently and accurately approximate UIR $\mathcal{R}^u$, determined by some prediction models, e.g., a classifier, for each of the $n$ subspaces, acting as $\{\mathcal{I}_i\}_{i \leq n}$.

\begin{figure}
\vspace{-5pt}
\small
    \centering
    \includegraphics[width=0.45\textwidth]{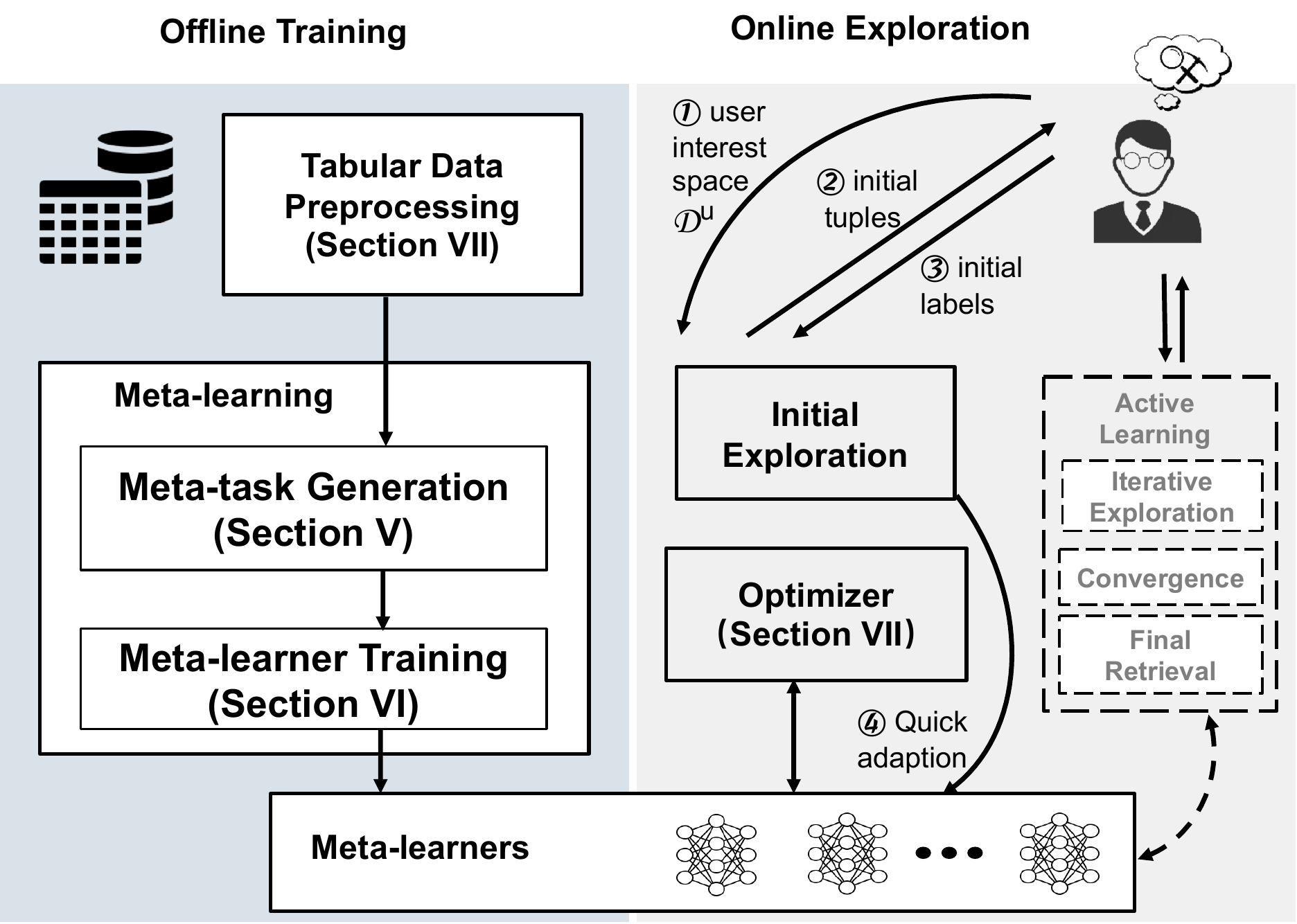}
\vspace{-5pt}
    \caption{Overview of Learn-to-explore Framework}
    \label{fig2}
\end{figure}
\vspace{-5pt}
\subsection{LTE Framework}
\label{subsec:framework}

A bird's-eye view of LTE framework is shown in Figure~\ref{fig2}.
It consists of four functional modules, {\it Meta-learning }, {\it Initial Exploration}, {\it Preprocessing}, and {\it Optimizer} modules, operating in
and two phases, \textit{offline training} and \textit{online exploration} phases.

\textit{\textbf{Meta-Learning}} is the core module of offline training phase. Its functionality is on training meta-learners by automatically generated meta-tasks.
As aforementioned, the data space of a database is decomposed into a set of subspaces, which we term {\it meta-subspaces}, $\{D^{M}_{i}\}$.
The components {\it meta-task generation} is in charge of generating meta-tasks, each of which contains a \emph{support set} and a \emph{query set} (Section~\ref{sec:taskgen}) of a meta-subspace. Then, the {\it meta-learner training} locally updates the meta-learner by support sets, and globally updates the meta-learner by query sets (Section~\ref{sec:meta-train}).


\textit{\textbf{Initial Exploration}} is the core module of online exploration phase. 
 At the initial stage of data exploration, a user first selects his/her interesting attributes  from the database schema to form a user interest space $\mathcal{D}^{u}$. The $\mathcal{D}^{u}$ is decomposed into a set of subspaces mapped to meta-subspaces.
 Then, he/she is presented with a selected set of initial tuples of UIS for labelling, the number of which is constrained by a given budget.
The selection of initial tuples is similar to the support set construction during meta-learning training (Section~\ref{sec:meta-train}).
Then, user labels are collected\footnote{As pointed in~\cite{10.14778/3275536.3275542}, collecting user's labelling feedback belongs to the field of human-computer interaction and is beyond the scope of this paper.}.
Finally, user labels are fed to pre-trained meta-learners on-the-fly to fast adapt to the real user interests.
The adapted meta-learners can determine the result UIR.

\textit{\textbf {Preprocessing}} is to convert an input tabular dataset into a series of composite vectors that can be fed to meta-learners, i.e., neural networks. The input of the module is a sampled database, achieving good feature representability and data scalability. The output of the module is feature-rich and high-dimensional vectors, conforming to the input of NN training. Details are reported in Section~\ref{sec:pre_opt}.

\textit{\textbf {Optimizer}} is heuristically dedicated to adjusting UIS predicted by each meta-learner in few-shot exploration. For each subspace, the module takes user labeled tuples during the initial exploration as input. After that, it takes two optimization steps for reducing false positives and false negatives, in order to polish the prediction results. Details are covered in Section~\ref{sec:pre_opt}.

\textit{\textbf {Other IDE Modules.}}
 Notice that our LTE framework can also be plugged to existing IDE systems~\cite{10.1145/2588555.2610523,7539596,10.14778/3275536.3275542} by connecting the trained meta-learners to active learning mechanisms.
For being self-contained, we also briefly review existing IDE modules~\cite{10.1145/2588555.2610523,7539596,10.14778/3275536.3275542} that can be combined with our LTE framework to make a complete system:
\textit{1) Iterative exploration.} If a user wants to continue exploring after the initial exploration phase, active learning can be employed to feed more labelled tuples to the meta-learner for further training~\cite{ren2020survey}. 
\textit{2) Convergence.} The user can set budgets for labelling, or use data visualization methods~\cite{visual,10.14778/2732951.2732964,10.14778/2735479.2735485,10.14778/2732240.2732250} to determine whether the exploration should be stopped. If such prerequisites are made, our framework can incorporate additional indicators (like \textit{three-set metric} in~\cite{10.14778/3275536.3275542}) for supporting the determination of exploration convergence.
\textit{3) Final retrieval.} An IDE system returns a sampled (or complete) set of user interest tuples, or infers corresponding query regions based on trained classifiers. The results can also be transformed to query filters (e.g., in SQL), if prerequisite assumptions about UIR and query templates are made~\cite{10.1145/2463664.2465220,10.1145/1559845.1559902,10.1145/2396761.2398507,10.14778/2367502.2367542,10.1145/2588555.2610523}.

\section{Meta-learning Process}
\label{sec:meta_over}
\begin{figure*}[ht]
\begin{center}
\vspace{-30pt}
\includegraphics[width=0.75\textwidth]{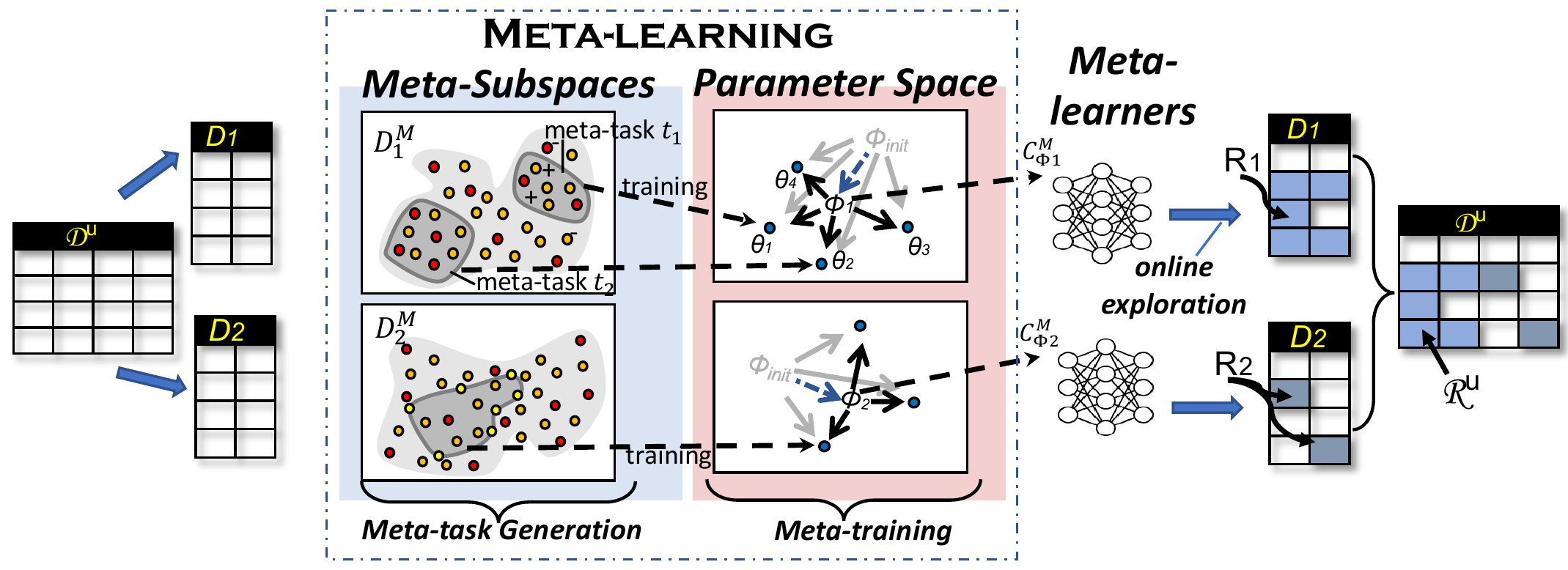}
\end{center}
\vspace{-10pt}
\caption{The Meta-learning Process}
\label{lp}
\end{figure*}



\vspace{-5pt}
\subsection{Concepts}

The core of the LTE framework is the meta-learning process, which is formalized as follows.
\begin{definition}[\textbf{Meta-learning Process}]
\label{def:ml}
Given a pre-defined meta-subspace $D^{M}$, a meta-task set $\mathcal{T}^{M}$, and a meta-learner $\mathcal{C}_{\theta}^{M}$ with randomly initialized parameter $\theta$, the meta-learning process can be modeled as a function $F_{M}$, as follows.
\begin{align}
F_{M}(\mathcal{T}^{M},\mathcal{C}_{\theta}^{M})\to \mathcal{C}_{\phi}^{M}
\end{align}
\end{definition}

Here, $\phi$ refers to an initialization parameter trained with meta-task set $\mathcal{T}^{M}$, and $\mathcal{C}_{\phi}^{M}$ refers to the learned meta-learner equipped with $\phi$. A quality $\phi$ helps $\mathcal{C}_{\phi}^{M}$ in efficiently approaching towards the optimization target during the online exploration phase. Next, we formalize the concept of a meta-task.

{\begin{definition}[\textbf{Meta-task}]
\label{def:mt}
A meta-task $t$ of a meta-task set $\mathcal{T}^{M}$ has three parts, a simulated UIS $R_t^M$, a support set $\mathcal{S}_{t}^{sp}$ and a query set $\mathcal{S}_{t}^{qs}$.
\vspace{-5pt}
\begin{align}
t:(R_t^M,\mathcal{S}_{t}^{sp},\mathcal{S}_{t}^{qs})
\end{align}
\end{definition}
}
Following typical meta-learning settings~\cite{hospedales2020metalearning,10.5555/3305381.3305498}, a meta-task is expected to be associated with a support set and a query set, whereas the support set is used for updating the meta-learner with local updates and the query set is used for updating the meta-learner with global updates.
Specifically, for data exploration, a meta-task $t$ is associated with a simulated UIS $R_t^M$, which is automatically generated. 

Then, of a meta-task $t$, both the corresponding the support set ($\mathcal{S}_{t}^{sp}$) and query set ($\mathcal{S}_{t}^{qs}$) consist of a certain number of labelled tuples, where a tuple is labelled by checking whether it is within the UIS. Differently, the support set is for simulating user actions of labeling during the exploration, and the query set is for simulating the evaluation of the trained meta-learner.


In Figure~\ref{lp}, $t_1$ is a meta-task on meta-subspace $D^M_1$, which consists of UIS (the shaded region), a support set (red points), and a query set (yellow points).
Note that meta-tasks of the same meta-subspace can share the tuples of the support and query sets. For example, meta-tasks $t_1$ and $t_2$ offer different coverages to the same set of red and yellow points.


\subsection{The Learning Process}

The meta-learning process can be viewed as a search optimization problem on the parameter space of the meta-learner, i.e., the domain of the parameter matrix of a neural network.
The parameter space is huge~\cite{hospedales2020metalearning}. It is known that a quality initialization parameter enables a fast convergence to the optimization target~\cite{10.5555/3305381.3305498} of the huge parameter space.

Intuitively, the initialization parameter is expected to have optimization distances uniformly close to the parameters corresponding to real tasks (during the online exploration). However, the challenges are two-fold: 1) the real tasks are of high diversity, which cannot be enumerated during the pre-training; 2) the parameter space is enormous, whereas conventional search optimization methods fall short. 

In our work, the former is addressed by automatically generated meta-tasks, consisting of a set of simulated UISs on a meta-subspace ({\it meta-task generation} in Section~\ref{sec:taskgen}). The latter is addressed by a meta-training algorithm under the ``gradient by gradient'' setting, taking quadratic derivation for the optimization of initialization parameters ({\it meta-training} in Section~\ref{sec:meta-train}).

An example of the meta-learning process is shown in Figure~\ref{lp}. Suppose a meta-learning process on meta-subspace $D^M_1$, which is to find an initialization parameter ($\phi_1$) from a random parameter ($\phi_{init}$).
It is expected that the optimization distances from $\phi_1$ to ``anchor parameters'' (e.g.,  $\theta_{1}\sim \theta_{4}$) are uniformly close.
An anchor parameter corresponds to a meta-task (e.g., $\theta_1$ corresponds to $t_1$), which can be obtained through the meta-training with its corresponding meta-task.
The functionality of an anchor parameter is to guide the direction of the search optimization of the meta-learning.
Thus, the quality of meta-learning process depends on the quality of meta-tasks.
Next, we discuss how the meta-tasks are generated.





\vspace{5pt}
\section{Meta-task generation}
\label{sec:taskgen}
\subsection{Meta-task Generation Algorithm}

In this section, we investigate how meta-tasks can be automatically generated for each meta-subspace,
so that meta-learning can be enabled in an unsupervised way.
In general, simulations require representing the key characteristics or behaviors of the data exploration process, so that two principles can be summarized for designing a meta-task.
\begin{itemize}
  \item {\it Faithfulness:} The tuples of support/query set should conform to the data distributions of a given meta-subspace, making a basis for the quality inference of UISs.
      Otherwise, the ``bias'' can be prorogated to the anchor parameters, thus affecting the optimization of initial parameters.
  \item {\it Generality:} A meta-task should be flexible in covering different tendencies of UISs, i.e., being general in putting together different pieces of data interests of a meta-subspace.
\end{itemize} 


For the first principle, a natural solution is to sample tuples conforming to the data distribution of a meta-subspace.
In this paper, we opt to a clustering-based sampling method, (e.g., $k$-means~\cite{loh2011classification}), which is proved to be primitive and effective for summarizing data insights~\cite{10.1145/2588555.2610523}.

For the second principle, we construct a general form of UIS, which can be represented as the composition of any set of convex parts on a meta-subspace, thus being general in supporting arbitrary shaped UISs,
according to the convex decomposition theory~\cite{Convex,Approximate_Convex}.

The overall process about meta-task generation is formalized in Algorithm~\ref{alg:taskgen}.
In general, the process consists of two steps, clustering step (Section~\ref{subsec:cluster}) and task generation step, where the latter contains UIS formulation (Section~\ref{subsec:taskgen}) and support/query set formulation (Section~\ref{subsec:setgen}).

\begin{algorithm}[ht]
\scriptsize
    \caption{Meta-task generation}
    \raggedright
    {\bf Input:} {Parameters $k_u$, $k_s$, $k_q$, $\alpha$ and $\psi$}\\
    {\bf Output:} a meta-task set $\mathcal{T}^M$\\
    \begin{algorithmic}[1]
    \STATE Perform three rounds of $k$-means clustering ($k$= $k_u$, $k_s$, and $k_q$), get cluster center sets $C^u$, $C^s$, and $C^q$, and calculate $P^u$ (Section~\ref{subsec:cluster});
    \STATE Generate UISs based on $C^u$, $P^u$, $\alpha$ and $\psi$ (Section~\ref{subsec:taskgen});
    \STATE Get Support/Query set on $C^s$, $C^q$ and UISs (Section~\ref{subsec:setgen});
    \STATE Collect UISs and corresponding support/query sets to get $\mathcal{T}^M$.
    \RETURN $\mathcal{T}^M$;
    \end{algorithmic}
    \label{alg:taskgen}
\end{algorithm}

\subsection{Clustering Step}\label{subsec:cluster}

Clusters (or cluster centers) can be viewed as a lightweight summary of a meta-subspace.
During the clustering step, we perform $k$-means clustering independently for three rounds\footnote{The clustering is run on a randomly sampled (1\%) subset of the tuples of the meta-subspace for scalability.}, because a meta-task $t$ is a triple, i.e., $t:(R_t^M,\mathcal{S}_{t}^{sp},\mathcal{S}_{t}^{qs})$. Each round is with a different parameter $k$ (i.e., $k_u$ for simulated UIS, $k_s$ for support set, and $k_q$ for query set).
Accordingly, we get three sets of cluster centers $C^u=\{c_i^u\}_{i \leq k_u}$, $C^s=\{c_i^s\}_{i \leq k_s}$, and $C^q=\{c_i^q\}_{i \leq k_q}$, based on which the simulated UIS, support set, and query set of a meta-task are generated.

During the clustering, we maintain two proximity matrices for the efficiency of subsequent steps, $P^u$ and $P^s$, based on $C^u$ and $C^s$.
The first matrix $P^{u}$ stores $k_{u} \times k_u$ elements, representing the distances between the $k_u$ cluster centers of $C^u$, for constructing the simulated UIS (Section~\ref{subsec:taskgen}). The second $P^{s}$ stores $k_{s} \times k_u$ elements, representing the distances between the $k_s$ cluster centers in $C^s$ and the $k_u$ cluster centers in $C^u$, used to expand the feature vectors of UIS (Section~\ref{subsec:clf}) and few-shot prediction optimization (Section~\ref{sec:pre_opt}). Without loosing generality, \textit{Euclidean distance} is employed for measuring the proximity. The proximity matrices can be done in $O(k^2_u+k_s\cdot k_u)$.

\subsection{UIS Formulation} \label{subsec:taskgen}

The generated meta-task set is expected to offer good coverage of UIS, which can be arbitrarily shaped in low-dimensional spaces. According to the convex decomposition theory, a region (or UIS) can be viewed as a combination of multiple intervals of different lengths in a 1D subspace or multiple 2D convex polygons in a 2D subspace, and similarly in higher dimensions.
Therefore, the UIS of a meta-task can be formulated by randomly combining a set of convex shaped parts on a meta-subspace. In this paper, we implement an efficient and straightforward generation method, which utilizes sampled tuples to construct multiple external convex regions to combine into the final simulated UIS. Moreover, we can control the size of a part and the number of parts, by changing the selection of the sampled tuples.

To generate a meta-task, we start by constructing a simulated UIS, which requires three steps. First, we randomly select a cluster center $c_j\in C^u$, and retrieve the set $S_j$ of $c_j$'s $\psi$ nearest neighbors (i.e., cluster centers), where $S_j=\psi NN(c_j)$ and  $S_j\subseteq C^u$. It can be done with the proximity matrix $P^u$ in $O(k_u)$.

Second, we build the \textit{convex hull} for $S_j$, represented by $Cvx(S_j)$, which is the largest circumscribed convex polygon for the cluster centers.
Notability, there can be other options for the circumscribed region, such as minimum bounding rectangles or circles. It can even be concave, as long as the selected cluster centers are circumscribed. In our implementation, convex hulls are adopted for their simplicity.
The convex hull serves as the basic building block of a UIS, which can be done in $O(\psi\cdot log(\psi))$.
The first two steps are repeated until $\alpha$ convex hulls are collected.

Finally, the $\alpha$ convex hulls are combined to get a simulated UIS, $R^M_t = \bigcup_{j\leq \alpha} Cvx(S_j)$. 
Notice that the UISs in existing works can be viewed as special cases generated by the above method. For example,~\cite{10.14778/3275536.3275542} assumes the UIS as a connected convex region ($\alpha = 1$).
In our implementation, we do not explicitly maintain the exact shapes of $R^M_i$. All we need is to determine, during the offline training, if a point of the meta-subspace is within the given UIS, which can be transformed into determining if a point is located within any of the $\alpha$ convex hulls. It can be done in $O(\alpha\cdot log(\psi))$. In empirical studies (Section~\ref{subsec:ger_exp}), we also consider different combination of $\psi$ and $\alpha$ as a UIS mode, and examine the performance on various modes.

\vspace{-5pt}
{\subsection{Support/Query Set Formulation}
\label{subsec:setgen}}

We can use a generated UIS to formulate the corresponding support and query sets. we first take the $k_{s}$ cluster centers from $C^s$ as tuples of the support set.
The label $y$ of each tuple is determined by checking if it belongs to the corresponding UIS. To increase the generality of meta-training, we further sample a few tuples randomly from the meta-subspace.
Therefore, the size of the final support set is $k_{s}+\Delta$\footnote{The default $\Delta$ is $5$ in our implementation.}. The query set is built in a similar way on $C^q$. The size of the query set is $k_{q}+\Delta$.

Notably, since the role of the support set is to simulate the set of tuples labeled by users, the initial tuples for online exploration in the LTE framework are also generated by the clustering step, for a subspace. Then, for each tuple of a subspace, a user needs to label it for initial exploration\footnote{If there exist assumptions on the UISs of subspaces~\cite{10.1145/2588555.2610523,7539596,10.14778/3275536.3275542}, we can put the subspaces that have a conjunctive relationship into a group to reduce the number of subspaces labeled by the user.}.


\vspace{5pt}
\subsection{Discussion}
\textit{\textbf{Dynamic Maintenance.}} 
The meta-learner is trained on the basis of meta-tasks, and meta-tasks are built on sampled tuples. So, one only needs to check if sampled tuples should be updated to decide if the meta-tasks and meta-learners should be updated, when the data distributions of the meta-subspaces change. Then, the problem is reduced to check for each subspace whether its corresponding clustering results violate, if the exploratory database is updated. The solution is to capture the locality of dynamic changes to data distributions of subspaces, corresponding sampled tuples set, and meta-tasks.
To this point, existing works of dynamic clustering~\cite{GanT17,GongZY17} can be applied. Details about dynamic clustering are beyond the scope of the paper are omitted due to page limits.


\textit{\textbf{Splitting Data Space to Meta-subspaces.}}
The data space should be split into a set of mutually exclusive meta-subspaces in the offline phase. One may establish as many meta-subspaces as possible,
for the matching with subspaces specified in the online phase.
However, one may need to generate $\binom{|A|}{d}$ $d$-dimensonal meta subspaces for covering all possibilities of splitting a $|A|$-dimensional space, which can be costly.
On one hand, we can determine some commonly used meta-subspaces based on the semantic/dependency relationship between attributes, or logs of user exploration. On the other hand, even if the meta-learner is not used for the subspace, the basic NN classifier combined with tabular data preprocessing still achieves better performance than existing methods (Section~\ref{subsec:ger_exp}).
In our implementation, the domain space is randomly split into meta-subspaces, because we assume zero knowledge about data semantics and user priors.

\section{Meta-learning training}
\label{sec:meta-train}
Given a meta-subspace ${D}^M$ with the meta-task set $\mathcal{T}^M$. Each meta-task $t\in \mathcal{T}^M$ contains a simulated UIS $R^M_t$, a support set $\mathcal{S}_{t}^{sp}$ and a query set $\mathcal{S}_{t}^{qs}$.
Both query and support sets are composed of a set of $2$-tuples $\{\tau,y_{_R,_\tau}\}$,
where $\tau\in D^{M}$ is a meta-subspace tuple, and $y_{_R,_\tau}$ is the label indicating whether $\tau$ belongs to UIS $R^M_t$ of $t$.
The meta-training goal is to find suitable initialization parameters $\phi$, so the neural network classifier $\mathcal{C}_{\theta}^M$ can fast adapt to $\mathcal{C}_{\phi}^M$.

We first introduce a UIS classifier based on neural networks in Section~\ref{subsec:clf}, and memory-augmented optimization for meta-learning in Section~\ref{subsec:mem}. After that, we propose the meta-learning algorithm in Section~\ref{subsec:train_alg}.


\subsection{Basic UIS Classifier}
\label{subsec:clf}
 We introduce a NN classifier, which contains three building blocks: \textit{UIS feature embedding block}, {\it Data tuple feature embedding block}, and {\it Classification block}.

\textbf{\textit{UIS Feature Embedding Block ($f_{\theta_{R}}$).}} To enrich the input features of the classifier in the few-shot exploration. We construct a $0/1$ vector of length $k_{s}$ from the set $C^s$, where each vector bit corresponds to a cluster center in $C^s$. The bit is assigned to $1$, if the user is interested in the corresponding cluster center, and $0$ otherwise. Notice that the bit position of the vector representing a cluster center is fixed and is therefore consistent all through the training phase. To a certain extent, the vector reflects  structural features of UIS for a given task. Meanwhile, $C^s$ as a predetermined unified set can ensure the comparability of different UISs' features, so that the UIS-Feature Embedding Block in Section~\ref{subsec:mem} can extract higher-level mode information from a large number of UISs' features.

Since $k_{s}$ reflects the number of tuples to be labelled initially with a limited budget, it corresponds to a small value.
As a result, the feature vectors in some fine-grained UIS may be highly sparse.
So, we enlarge the vector from set $C^s$ to set $C^u$, with a heuristic expansion technique.

Specifically, for any bits of the original $k_{s}$-bit feature vector are $1$, we first retrieve the cluster centers represented by such bits in $C^s$, then get their $l$-nearest neighbors from  $C^u$, by using the precomputed $k_s \times k_u$ proximity matrix $P^s$.
$l$ represents the degree of heuristic expansion, which is set to a constant value (e.g., $l=0.1\times k_{u}$ by default). Finally, we redefine a $0/1$ vector of length $k_{u}$ and set all the bits corresponding to the cluster centers located in $C^u$ to $1$.
For a meta-task $t$'s UIS $R^M_t$, this vector is known as the UIS feature vector $v_{R}\in \mathbb{R}^{k_{u}}$. Thus, our embedding block $f_{\theta_{R}}$ can be expressed as:
\begin{align}
 \label{eq1}
emb_{R} = f_{\theta_{R}}(v_{R})
\end{align}
where $\theta_{R}$ represents the parameters of fully connected layers, and $emb_{R}$ represents the output of the embedding layer.

\textbf{\textit{Data Tuple Embedding Block ($f_{\theta_{\tau}}$).}} Assuming that a data tuple $\tau$ 's representation vector is of size $N_{r}$, denoted by $v_{\tau}\in \mathbb{R}^{N_{r}}$, this block can be written as:
\begin{align}
 \label{eq2}
emb_{\tau} = f_{\theta_{\tau}}(v_{\tau})
\end{align}
where $\theta_{\tau}$ represents the fully connected layer parameters, and $emb_{\tau}$ represents the output of the embedding layer.
For $f_{\theta_{R}}$ and $f_{\theta_{\tau}}$, we set the embedding size to $N_{e}$. Thus, $emb_{R}$ and $emb_{\tau}$ are equally sized, $emb_{R}$, $emb_{\tau}\in \mathbb{R}^{N_{e}}$.

\textbf{\textit{Classification Block ($f_{\theta_{clf}}$).}} Given a meta-task $t$'s UIS
 feature embedding $emb_{R}$, and a list of the corresponding data tuple embeddings $emb_{\tau}$ for $\tau\in \mathcal{S}_{t}^{sp}$ or $\mathcal{S}_{t}^{qs}$, we can get the predicted label $\hat{y}_{_R,_\tau}$ by classification block $f_{\theta_{clf}}$:
 \begin{equation}
  \label{eq3}
 \hat{y}_{_{R},_\tau} =  f_{\theta_{clf}}([emb_{R},emb_{\tau}])
 \end{equation}
where $[emb_{R},emb_{\tau}]$ is the concatenation of the UIS embedding and the data tuple embedding, and $\theta_{clf}$ denotes the parameters of fully connected layers for classification block.

  Thus, the parameters $\theta$ for $\mathcal{C}_{\theta}^M$ is $\{\theta_{R},\theta_{\tau},\theta_{clf}\}$, and the goal of meta-learning is to get the learned initialization parameters $\phi=\{\phi_{R},\phi_{\tau}$ $,\phi_{clf}\}$.

\subsection{Memory-Augmented Optimization.}
\label{subsec:mem}
Inspired by~\cite{lee2019melu,dong2020mamo,graves2014neural}, we utilize extra memories (parameters matrices) to store and update some model parameters to overcome the problem that the conventional meta-learning method is easy to slip into the local optimum. Since the basic meta-learning method assigns the same learned initialization parameters (e.g., $\{\phi_{R},\phi_{\tau},\phi_{clf}\}$) to model parameters (e.g. $\{\theta_{R},\theta_{\tau},\theta_{clf}\}$) for all tasks during meta-training and the actual use, we hope these learned parameters can be fine-tuned appropriately on different tasks to obtain task-wise parameters.
Based on these initialization parameters, we can use labeled tuples to train the classifier more efficiently in the optimization direction of the current task. Thus, we introduce two types of memories similar to~\cite{dong2020mamo}, \textit{UIS-feature memory} and {\it embedding-conversion memory}. The former memory focuses on adjusting the learned initialization parameters of the UIS feature embedding block. The latter memory focuses on the conversion of parameters before inputting $[emb_{R},emb_{\tau}]$ into $f_{\theta_{clf}}$.
The two memories will be updated simultaneously with the meta-learning process. Note that the LTE framework is orthogonal to all existing MAML-based meta-learning algorithms.

\textbf{\textit{UIS-Feature Memory.}} The UIS-feature memory includes the UIS embedding parameters matrix $M_{R}$, and the UIS feature vector matrix $M_{v_{R}}$. Given a certain task $t$, the meta-learned initialization parameters of the UIS feature embedding block is $\phi_{R}$. Our goal is to fine-tune $\phi_{R}$ to obtain the task-wise initialization parameter $\theta_{R}$:
\begin{equation}
\label{eq4}
\theta_{R} \Leftarrow \phi_{R}-\sigma  \omega_{R}
\end{equation}
where $\omega_{R}$ represents the parameters that need to be adjusted on $\phi_{R}$ (i.e., $\omega_{R}$ is a bias term~\cite{lee2019melu, dong2020mamo}), and $\sigma\in[0,1]$ is a hyper-parameter  that indicates how much $\phi_{R}$  needs to be updated. Since we expect $\omega_{R}$ to be associated with a specific task, we adopt the following method to obtain it:

First, we calculate an attention value $a_{R}$ form  $M_{v_R}$ that stores information relevant to a UIS feature vector $v_{R}$:
\begin{equation}
 \label{eq5}
a_{R} = Sim(v_{R},M_{v_R})
\end{equation}
where $M_{v_R}\in \mathbb{R}^{m\times k_{u}}$ is a $m\times k_{u}$ matrix storing the mode information extracted from the UIS feature vectors of all meta-tasks during the meta-learning training. Here, $m$ is a hyper-parameter, representing the number of implicit modes/patterns~\cite{lee2019melu, dong2020mamo} that we want to extract from the UIS feature vectors of the meta-task set. $Sim$ function calculates the \textit{Cosine similarity} between a UIS feature vector $v_{R}$ and $M_{v_R}$, which is normalized by $SoftMax$ function. Thus, we can get $a_{R} \in \mathbb{R}^{m}$. Then, the retrieval attention value $a_{R}$ is applied for extracting  parameters $\omega_{R}$ from  the memory $M_{R}$:
 \begin{equation}
  \label{eq6}
 \omega_{R}=a_{R}^{\mathrm{T}}M_{R}
 \end{equation}
where each row of $M_{R}$ keeps the parameters of the UIS feature embedding block. Since the UIS embedding block may be comprised of more than one layer and more than one parameter, $M_{R}\in \mathbb{R}^{m\times |\theta_{R}|}$ stores all the parameters in the same form as the parameters in the UIS embedding block (parameters size is $|\theta_{R}|$ for both).

The two memory matrices are randomly initialized at the training beginning and will be updated during global update phase of meta-learning.

\textbf{\textit{Embedding-Conversion Memory.}} The memory aims to obtain task-wise embedding conversion parameters for  $[emb_{R},emb_{\tau}]\in \mathbb{R}^{2N_{e}}$, corresponding to a specific task. We employ an extra parameters matrix $M_{cp}\in \mathbb{R}^{N_{e}\times2N_{e}}$ to store the conversion parameters. Thus, Equation~\ref{eq3} can be rewritten as :
 \begin{equation}
   \label{eq7}
 \hat{y}_{_R,_\tau} =  f_{\theta_{clf}}([emb_{R},emb_{\tau}])=f_{\theta_{clf}}(M_{cp}\cdot[emb_{R},emb_{\tau}])
 \end{equation}
Similar to the UIS-feature memory, we employ the attention value $a_{R}$ to retrieve the parameters $M_{cp}$ from the global conversion parameters matrix $M_{CP}$:
 \begin{equation}
  \label{eq8}
M_{cp}= a_{R}^{\mathrm{T}} \cdot M_{CP}
\end{equation}
where $M_{CP}\in \mathbb{R}^{m\times N_{e}\times2N_{e}}$ stores the ``equalization" conversion parameters with $m$ implicit modes/patterns, which are extracted from the conversion parameters obtained on all meta-tasks.
During meta-learning training, $M_{cp}$ will be updated in the local update phase together with the updates of parameters of the classifier, and $M_{CP}$ will be updated with $M_{cp}$ in the global update phase.


\subsection{Training strategy}
\label{subsec:train_alg}
Algorithm~\ref{alg:meta_train} depicts the entire meta-learning process. At the beginning of the training, we randomly initialize all global parameters (including the parameters of the classifier and extra memories): $\phi_{R},\phi_{\tau},\phi_{clf},M_{R},M_{v_R},$ and $M_{CP}$.
After that, according to the sequence of parameter updates, we divide the training process into \textit{local} and \textit{global} phases.

\begin{algorithm}[t]
\scriptsize
    \caption{Training process of meta-learning}
    \raggedright
    {\bf Input:} {Meta-task set $\mathcal{T}^{M}$; UIS feature vector $v_{R}$ for $t\in \mathcal{T}^{M}$, Representation vector $v_{\tau}$ and true label $y_{_R,_{\tau}}$ for tuple $\tau\in \mathcal{S}_{t}^{Sp}$, $\mathcal{S}_{t}^{Qs}$; Hyper-parameters $\eta,\beta,\gamma,\sigma,\rho,\lambda$};\\
    {\bf Output:} {Learned parameters: $\phi_{R},\phi_{\tau},\phi_{clf},M_{R},M_{v_R},M_{CP}$;}\\
	\begin{algorithmic}[1]
    \STATE Random initialize $\phi_{R},\phi_{\tau},\phi_{clf},M_{R},M_{v_R},M_{CP}$;
    \WHILE{not reach training epochs}
    \FOR{$t\in \mathcal{T}^{M}$}
    \STATE \underline{Get $a_{R}$, $\omega_{R}$~(Equation 7, 8), Initialize $\theta_{R}$~(Equation~\ref{eq4}});
    \STATE \underline{Initialize $M_{cp}$~(Equation 10), $\theta_{\tau},\theta_{clf}$~(Equation~\ref{eq9})};
	\FOR{$\{\tau,y_{_R,_{\tau}}\}\in \mathcal{S}_{t}^{sp}$}
	\STATE \underline{Get $emb_{R},emb_{\tau}$~(Equation~\ref{eq1}, \ref{eq2}});
	\STATE \underline{Get prediction label $\hat{y}_{_R,_{\tau}}$~(Equation~\ref{eq7}});
	\STATE \underline{Locally update  $\theta_{R},\theta_{\tau},\theta_{clf}$~(Equation~\ref{eq10}});
	\STATE \underline{Locally update $M_{cp}$ by back-propagation};
	\ENDFOR
    \ENDFOR
    \STATE Globally update $M_{v_R},M_{R},M_{CP}$~(Equation~\ref{eq12}, \ref{eq13}, \ref{eq14});
 
    \STATE Globally update $\phi_{R},\phi_{\tau},\phi_{clf}$~(Equation~\ref{eq11});

    \ENDWHILE
    \RETURN $\phi_{R},\phi_{\tau},\phi_{clf},M_{R},M_{v_R},M_{CP}$;
    \end{algorithmic}
    \label{alg:meta_train}
\end{algorithm}

\textbf{\textit{Local Update on the Support Sets.}} The phase refers to the updating of local parameters: $\theta_{R},\theta_{\tau},\theta_{clf},M_{cp}$ on the support set. For each task $t\in \mathcal{T}^M$, we have support set  $\mathcal{S}_{t}^{sp}$. During the local update phase, we first initialize the local classifier parameters :
$\{\theta_{t},\theta_{d},\theta_{clf}$ $,M_{cp}\}$. We use Equation~\ref{eq4} for initialization of $\theta_{R}$, and Equation~\ref{eq10} for $M_{cp}$. Since $\theta_{\tau}$ and $\theta_{clf}$ do not involve memory-augmented optimization, we use the conventional meta-learning initialization method~\cite{10.5555/3305381.3305498}:
\begin{equation}
 \label{eq9}
\theta_{\tau}; \Leftarrow \phi_{\tau}; \theta_{clf} \Leftarrow \phi_{clf}
\end{equation}

The optimization goal for a single task in local training is to minimize the loss of the classification. Thus, the local parameters will be updated as:
\begin{equation}
 \label{eq10}
\theta_{*}\Leftarrow \theta_{*} - \rho\cdot\nabla_{\theta_{*}}LossFunc(y_{_R,_\tau},\hat{y}_{_R,_\tau})~ or~\nabla M_{cp}
\end{equation}
where $*$ could be any element in $\{R, \tau, clf\}$; $\rho$ is the learning rate for updating local parameters. It is worth noting that the parameters in $M_{cp}$ are also updated through back-propagation~\cite{graves2014neural}.

\textbf{\textit{Global Update on the Query Sets.}} The phase  aims to update global parameters: $\phi_{R},\phi_{\tau},\phi_{clf},M_{R},M_{v_R},$ and $M_{CP}$.

According to the ``gradient by gradient" setting of meta-learning training~\cite{10.5555/3305381.3305498}, we need to perform gradient descent on the locally updated gradient on the support set by minimizing the loss on the query set $\mathcal{S}_{t}^{qs}$ to update the global parameters $\phi_{R},\phi_{\tau},$ and $\phi_{clf}$. In order to save the cost of training, after the local update on support sets of all meta-tasks, we update the global parameters by taking one-step gradient descent like~\cite{dong2020mamo}. Thus, the global parameters are updated by
\begin{equation}
 \label{eq11}
\phi_{*}\Leftarrow \phi_{*} - \lambda \sum_{\mathcal{T}^M}\sum_{\mathcal{S}_{t}^{qs}}\nabla LossFunc(\hat{\theta_{*}})
\end{equation}
where $*$ could be any element in $\{R, \tau, clf\}$. $\hat{\theta}_{*}$ is the parameters of $\mathcal{C}_{\theta}^M$ after training on support sets, and $\lambda$ is learning rate.

In addition, $M_{R},M_{v_R},$ and $M_{CP}$ will also be updated as follows.  $M_{v_R}$ and $M_{R}$  will be updated as~\cite{dong2020mamo}:
 \begin{equation}
 \vspace{-5pt}
  \label{eq12}
 M_{v_R}=\eta\cdot(a_{R}\times v_{R}^{\mathrm{T}})+(1-\eta)M_{v_R}
 \end{equation}
where $\times$ denotes the cross-product, and $\eta$ is a hyper-parameter to control how much new UIS feature information is added. To ensure that the new information is added to the memory attentively, the attention value $a_{R}$ also be used. Similarly, the $M_{R}$ will be updated by
 \begin{equation}
  \label{eq13}
M_{R}=\beta\cdot(a_{R}\nabla_{\theta_{R}}(LossFunc(y_{_R,_\tau},\hat{y}_{_R,_\tau}))+ (1-\beta)M_{R})
 \end{equation}
where $\beta$ is the hyper-parameter to control how much new information is kept. $M_{CP}$ will be updated in the following~\cite{graves2014neural}:
 \begin{equation}
  \label{eq14}
M_{CP}=\gamma\cdot(a_{R}\otimes M_{cp}) +(1-\gamma)M_{CP}
\end{equation}
where $\otimes$ denotes the tensor product, and $\gamma$ is a hyper-parameter to control how much new information from conversion parameters should be added.

In the online exploration phase, the steps to train the meta-learners by user-labeled tuples are similar to the local update of meta-learning (see the underlined steps in Algorithm~\ref{alg:meta_train}), except that we directly use the learned global parameters and extra memories.

\textbf{\textit{Discussion.}} The overhead of the meta-learning is mostly dependent on the size of the meta-task set $|\mathcal{T}^{M}|$ (see Section~\ref{subsec:anl}). Since a well established $\mathcal{T}^{M}$ needs to traverse as many instances as possible in the meta-subspace, $|\mathcal{T}^{M}| \propto dim(\mathcal{D}^{M})$.
In summary, we apply meta-learning to low-dimensional subspaces, which, 1) conforms to the strategy of high-dimensional space decomposition of existing IDEs, and 2) significantly reduces the training overhead.

\section{Preprocessing and Optimization}
\label{sec:pre_opt}



\subsection{Tabular Data Preprocessing}
 It refers to the preprocessing for tabular before the meta-learner training, during the offline phase. A straightforward way is to use the maximum and minimum normalization to process the database (numerical attribute) tuples. However, the method is far from providing feature representations that guarantee the essential performance of NN classifiers. Moreover, training NN classifiers with simple normalization in a low-dimensional data space may cause gradient saturation~\cite{fan13relational},
 when training with few-labeled tuples.


In our implementation, the tabular representation is based on multi-modal attribute features, extracted by Gaussian mixture model (GMM)~\cite{reynolds2009gaussian} and Jenks natural breaks classification (JKC)~\cite{jenks1971error,jenks1977optimal}. The representation vector of an attribute value is the concatenation of two parts. The first part is a one-hot vector, indicating which GMM component or JKC interval the value belongs to. The second part is a value of range $[0,1]$, which is the normalized value on its corresponding GMM component or JKC interval.

Thus, given a database attribute $a_{j}$, and a set of tuples of $a_j$ sampled from the database $T_{j}^{s}$, the tabular representation transforms $\tau_j$ to $v_{\tau_j}$ (Algorithm \ref{alg:rep}). A tuple contains multiple attributes, and therefore the vector representation of the tuple is obtained by concatenating the vector representations on all attributes. Next, we discuss GMM and JKC models.

\textit{\textbf{Gaussian Mixture Model.}}
 We can capture the feature of the attribute values by training the GMM composed of a series of Gaussian distribution components. Suppose a batch of sample data for a certain attribute, we may get the set of Gaussian distribution components
$\{g_i\}_{i \leq |g|}$, in accordance to the specified number of components, $|g|$, for the GMM training. The mean and variance of each component $g_{i}$ are represented by $\mu_{i}$ and $\theta_{i}$, respectively. Then, given an attribute value $\tau_{j}$, we can compute the probability distribution $\{p_{1}^{j},..,p_{|g|}^{j}\}$ that the value belongs to the components, and choose the one maximizing the likelihood, as the GMM component corresponding to $\tau_{j}$.

\textit{\textbf{Jenks Natural Breaks Classification.}} JKC is a data clustering method designed to determine values' best arrangement to different classes, called Jenks Natural Breaks intervals, or JKC intervals. JKC divides the distribution of a numerical attribute into approximately smooth JKC intervals, $\{b_{i}\}_{i\leq |b|}$, by minimizing the variance within an interval, and maximizing the variance between different intervals. We can also specify the number of JKC intervals, $|b|$. Then, given an attribute value $\tau_{j}$, we can quickly determine which JKC interval it belongs to, by comparing with boundary values of different JKC intervals.

Notice that GMM is suitable for processing numerical attributes with distribution composed of one or more peaks, i.e., unimodal and multimodal distributions, according to\cite{xu2019modeling}.
Also, we find that there are a large number of numerical attributes with distributions composed of smooth intervals, like trends or time series, which are more suitable for being processed by JKC or other interval scanning techniques\cite{xie2014maximum}.
In addition, Using GMM or JKC for the entire exploratory database can be costly, though it is effective. A practical solution is to make tabular representation on sampled data, so that the scalability can be ensured.
 In our work, we use random sampling and limit the sampling ratio under 1\%.

\begin{algorithm}[t]
\scriptsize
    \caption{Tabular data preprocessing}
    \raggedright
    {\bf Input:} $T_{j}^s$, a sampled tuple set on attribute $a_j$; $\tau_j$, a tuple to be represented on $a_j$; 
    $|g|$, \# of GMM components; $|b|$, \# of JKC intervals;\\
    {\bf Output:} $v_{\tau_j}$, a represented vector for $\tau_j$\;\\
    \begin{algorithmic}[1]
    \STATE $\{g_i\}_{i \leq |g|}\gets$ \textit{\textbf{GMM($T_{j}^s$, $|g|$)}} or $\{b_i\}_{i \leq |b|}\gets$ \textit{\textbf{JKC($T_{j}^s$, $|b|$)}};
    \IF{using GMM}
    \STATE Compute $\{p_{1}^{j},..,p_{|g|}^{j}\}$ for $\tau_{j}$ and $k=argmax_{\kappa}$ $p_{\kappa}^{j}$ ;
   	\STATE $label_{j}\gets$\textit{\textbf{ont-hot($g_{k}$, $\{g_{1},..,g_{|g|}\}$)}};
 $Norm_{j}\gets\frac{\tau_{j}-\mu_{k}}{2\theta_{k}}$;
 \ENDIF
    \IF{using JKC}
	\STATE Find the JKC interval $b_{k}$ corresponding to $\tau_{j}$;
    \STATE $label_{j}\gets$\textit{\textbf{one-hot($b_{k}$, $\{b_{1},..,b_{|b|}\}$)}}; $Norm_{j}\gets\frac{\tau_{j}-b_{k}.{\bf min}}{b_{k}.{\bf max}-b_{k}.{\bf min}}$;
      \ENDIF
    \RETURN $v_{\tau_j} \leftarrow label_{j}\oplus[Norm_{j}]$;
    \end{algorithmic}
    \label{alg:rep}
\end{algorithm}

\subsection{Optimization for Few-shot Prediction.}
The purpose of a classifier is to predict a UIS during online exploration.
The quality of prediction can be measured by ``false positive" (FP) and ``false negative" (FN).
FPs refer to errors of falsely predicting tuples that are ``not interest'' as ``interest'' to a user.
FNs refer to errors of falsely predicting tuples that are ``interest'' as ``not interest'' to a user.
Given a well-trained classifier, the quality is much dependent on the ``few shots'' during the online training. we study optimizations for quality refinement.
Note that since we study UIS of arbitrary shapes, the optimization method is heuristic and preliminary.
It thus creates a vacuum for further optimization over specified UISs based on our approach.


\textbf{\textit{For FP Errors.}} We study an inevitable source of FP errors which are common for few-shot learning. 
An uninteresting tuples far away from user labeled tuples could be randomly predicted by a classifier as ``interesting", because of lacking of sufficient information. Thus, we hope to get a superset of UIS to fix such errors. First, we use positively labelled tuples by users as ``anchor points".
Then, we retrieve the proximate tuples of anchor points to build a set of large-scale circumscribed regions.
The combination of large-scale regions (called ``outer-subregion") is conceived to cover the real UIS.
For the tuples located within the outer-subregion, we follow the prediction of classifiers, so that there is no chance of a negative tuple being recognized as positive, and vice versa.
Also, classifiers revise a tuple from positive to negative, if the tuple is not covered by the outer-subregion, i.e., beyond UIS.




In our implementation, the approach of searching neighboring tuples is similar to expanding the UIS feature vector in Section~\ref{subsec:clf}. After tuples are labelled in the initial exploration phase, for each cluster center that is identified as ``interesting" in $C^s$, we search for $N_{sup}$ more proximate cluster centers from other cluster center set (e.g., $C^u$) by proximity matrix $P^s$(see Section~\ref{subsec:cluster}).
Parameter $N_{sup}$ reflects the extent of expansion, in our implementation, we set it to $k_u$ or $k_q$ multiplied by a certain scale factor. Then, we construct a circumscribed region, e.g., convex hull, on these cluster centers. Finally, all convex hulls are combined as the outer-subregion.

\textbf{\textit{For FN Errors.}}
Similarly, an inevitable source of FN errors originates in the randomness of classifier prediction under few-labeled tuples.
A type of FN errors appear as some small ``false'' regions within the real UIS, which are predicted as ``not interesting'' by classifiers.
As a result, we build a small-scale circumscribed region set (called \textit{inner-subregion}) from the positive tuples in the initial tuple set, which can be considered a subset of the UIS. We can infer that the tuples in the inner-subregion should be positive, since they are located within the UIS. Also, classifiers revise a tuple to positive, if it is falsely recognized as negative, i.e., located within an inner region.
The approach of constructing the inner-subregion is the same as that of the outer-subregion, except that the expansion step size $N_{sub}$ is significantly less than $N_{sup}$, which we term as conservative expansion.

\section{Results}
 \label{sec:ret}

 \subsection{Setup}
 \label{sec:exp_set}

 \textit{\textbf{Datasets.}} We use 2 public datasets, SDSS~\footnote{https://www.sdss.org/dr17/} and CAR~\footnote{https://data.world/data-society/used-CARs-data}, which are commonly used in previous works.
 SDSS is a scientific dataset of sky objects~\cite{Interactively}.
 We use $100$K tuples of $8$ attributes follow the settings of~\cite{10.14778/3275536.3275542}. CAR has $50$K tuples for second-hand car information in eBay. We select $5$ commonly used attributes out of $19$ attributes based on the guidance in~\cite{Interactively}.
 For each dataset, data space is randomly split into a set of 2D subspaces, in consistency with settings of baselines for fair comparison.

 \textit{\textbf{Baselines.}}
There are two state-of-the-arts for UIR/UIS classification, DSM~\cite{10.14778/3275536.3275542} and AL-SVM~\cite{7539596}.
The settings of DSM and AL-SVM follows the settings in ~\cite{10.14778/3275536.3275542} and \cite{7539596}, respectively.
AL-SVM uses active learning to select training tuples for SVM.
DSM improves AL-SVM by incorporating the polytope-based optimization.
To fully examine the performance of LTE, we list several variants, {\it Basic}, {\it Meta}, and {\it Meta*}. Basic uses basic UIS Classifiers without any optimization. Meta improves Basic with meta-learning.
Meta* adopts all optimizations proposed (i.e., using the optimizer module based on Meta), Note that the effect of the optimizer is completely dependent on the underlying results of the meta-learner, and the optimizer cannot be used alone.

 \textit{\textbf{Metrics.}} Accuracy and efficiency are two common metrics for IDE evaluation. Accuracy is measured by
 F1-score = $\frac{2 \cdot \mathrm{precision} \cdot \mathrm{recall}}{\mathrm{precision} + \mathrm{recall}}$.
 Efficiency refers to cost efficiency, which is constrained by budget $B$, i.e., the number of labelled tuples needed.

 \textit{\textbf{Parameters\footnote{The default parameters are bolded.}.}} \underline{\textit{Meta-task generation}}: we set the $k_{u}=100$ and $k_{q}=200$  in each meta-subspace. Since the size of support sets ($k_{s}+\Delta$) reflects the exploration budget $B$ for labelling, we trained the corresponding meta-learners under $B=\{{\bf 30}, 40, 50, 100\}$.
 We set $\alpha=1$ and $\psi=50$ for generating meta-tasks following the setting of baselines in Section~\ref{subsec:CompBase}. We set $\alpha=4$ and $\psi=20$ for generating meta-tasks to study the adaptability of our method to various types of complex UISs in Sections~\ref{subsec:ger_exp}.
  We also test the performance by varying the size of the meta-task set $|\mathcal{T}^M|$ as $\{1000,5000,10000,15000,{\bf20000}\}$.
 \underline{\textit{Meta-learning training}}: for the meta-learner, we set the embedding size $N_{e}=100$ and use the $Relu$ activation function between all layers. By searching for meta-learning training hyper-parameters~\cite{dong2020mamo}, we search $\eta,\beta,\gamma,\sigma,\rho,\lambda$ in $\{$$0.01$, $0.001$, $0.0001$, ${\bf 0.00005}\}$ and $m$ in $\{2,4,{\bf 6}\}$. The number of training epochs is in $\{1,2,3,{\bf 4}\}$. The training batch size is in $\{5,10,{\bf 15}\}$ and the training step size is in $\{5,10,20,{\bf 30}\}$ for the local update phase. \underline{\textit{Optimizer}}: we set $N_{sup}$ as $\{20\%,{\bf 30\%},40\%\}$ of $k_{u}$ and $N_{sub}$ as $\{{\bf 5\%},10\%,15\%\}$ of $k_{u}$.

\subsection{Comparison with Baselines}
\label{subsec:CompBase}

\begin{figure}
\subfigure[Accuracy w.r.t Dimension] {\label{subf:accvsdim}\includegraphics[width=0.45\columnwidth]{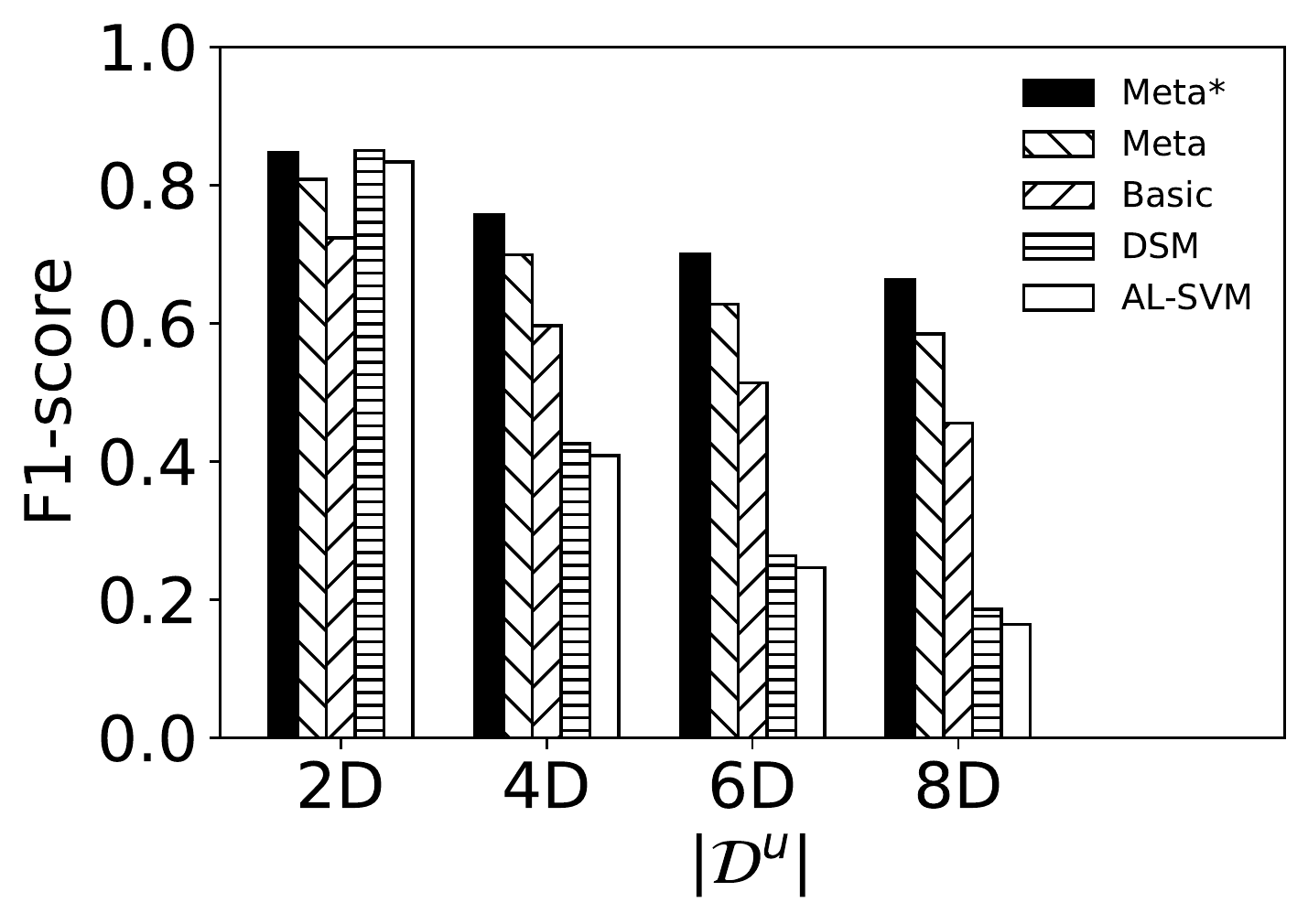}}
\subfigure[Efficiency w.r.t Dimension] {\label{subf:effvsdim}\includegraphics[width=0.45\columnwidth]{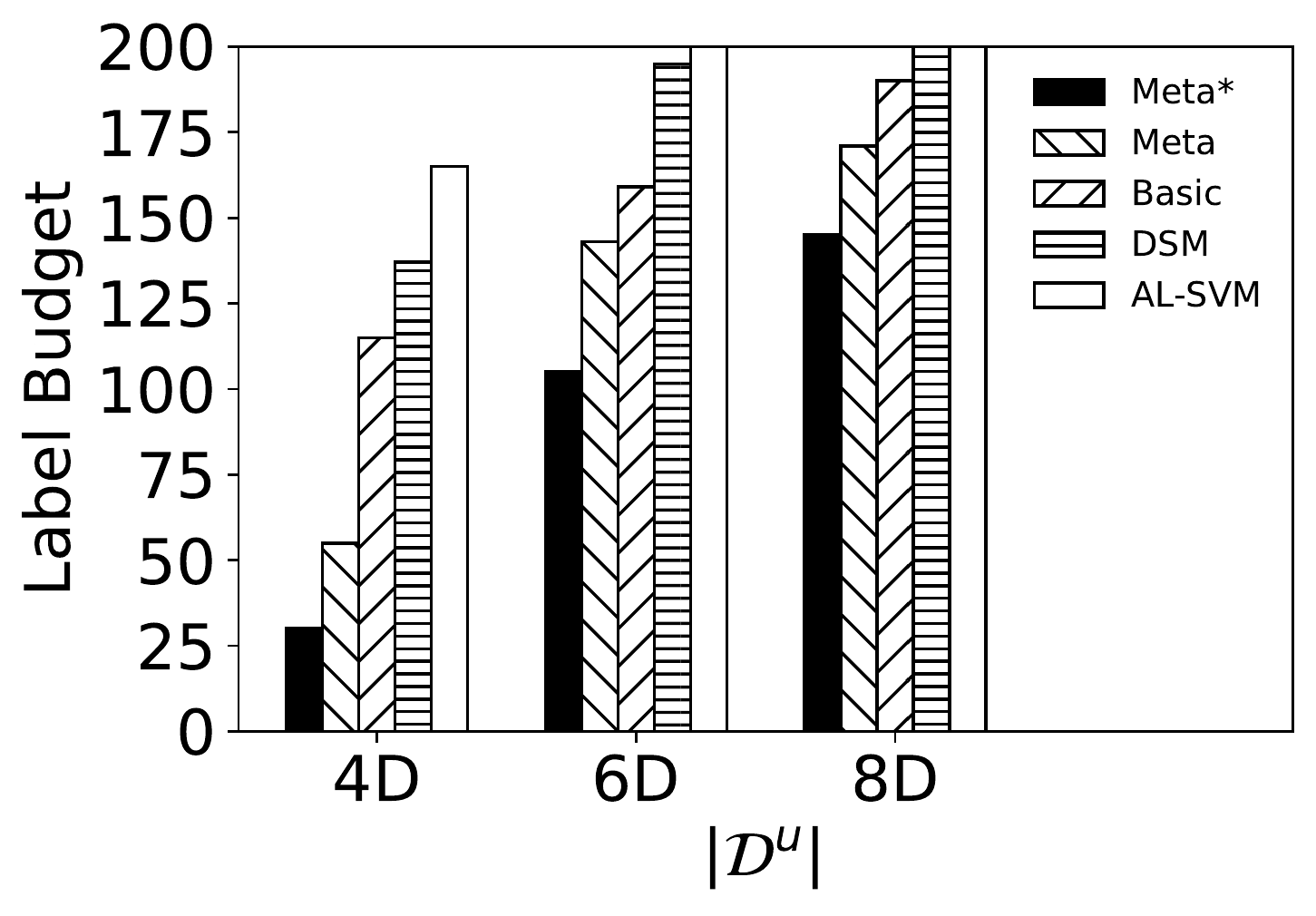}}
\vspace{-5pt}
\caption{Learn-to-explore vs. Baselines (SDSS)}
\label{fig:baseline}
\vspace{-10pt}
\end{figure}

For comparison with baselines, we follow DSM and AL-SVM's assumption on subspatial convexity and conjunctivity. Therefore, for a testing UIR in the high dimensional space, we construct convex UIS in  low-dimensional subspaces by {\it convex hull} model (i.e., fixing $\alpha=1$ and vary $\phi \in \{20,15,10,5\}$) and use the conjunctive property to unite these UISs to get final UIR. We totally generate 2,500 UIRs from 2D to 8D for testing, and the result is in Figure~\ref{fig:baseline}.
In addition, the result of our proposal without UIR assumptions is to be shown in Section~\ref{subsec:ger_exp}.
Notice that the cost of initial sampling~\cite{liu2018analysis} for baselines is not counted in their statistics, in all testings.

Figure~\ref{subf:accvsdim} examines the effect of dimensionality over the accuracy, by fixing $B$ to $30$.
It shows that the accuracy decreases w.r.t. the increase of dimensionality, for all competitors. It is because that the selection of representative tuples becomes more difficult, due to the sparse data distribution of a higher dimensional space.
Compared to the sharp drop of SVM-based methods, DSM and AL-SVM, the performance of NN-based methods are much more stable. In particular, the F1-score of DSM decreases about $75\%$ when the number of dimensions changes from $2$ to $8$. In comparison, the drop rate of all NN-based classifiers are steadily below $40\%$. Among them, the number of Meta* only drops about $18\%$, showing good scalability with dimensions.

Figure~\ref{subf:effvsdim} examines the effect of dimensionality over efficiency, by fixing F1-score to $0.75$. It shows that Meta* can achieve a given F1-score with a budget of less than 150 labeled tuples on 4-8D. However, DSM and AL-SVM require more than 150 labeled tuples in 6-8D (especially in 8D, far exceeding 150), which can be tedious for users. Since DSM outperforms AL-SVM in all testings, we only show the result of DSM in following experiments.

\begin{table}[h]
\vspace{-5pt}
\scriptsize
\caption{Accuracy w.r.t. UIS Modes (B=30)}
\vspace{-5pt}
\label{tab:acc_typ}
\setlength{\tabcolsep}{0.018\columnwidth}
\centering
\begin{tabular}{clccccccc}
\hline
\textbf{} & \multicolumn{1}{c}{\textbf{}} & \textbf{M1} & \textbf{M2} & \textbf{M3} & \textbf{M4} & \textbf{M5} & \textbf{M6} & \textbf{M7} \\ \hline
\multirow{5}{*}{\textbf{CAR}} & \multicolumn{1}{c}{\textbf{Meta*}} & 0.839 & 0.723 & 0.544 & 0.307 & 0.71 & 0.749 & 0.786 \\ \cline{2-9}
 & \textbf{Meta} & 0.795 & 0.667 & 0.486 & 0.266 & 0.606 & 0.673 & 0.731 \\ \cline{2-9}
 & \textbf{Basic} & 0.737 & 0.612 & 0.421 & 0.231 & 0.462 & 0.568 & 0.652 \\ \cline{2-9}
 & \textbf{SVM$^r$} & 0.712 & 0.562 & 0.331 & 0.127 & 0.450 & 0.531 & 0.624 \\ \cline{2-9}
 & \textbf{SVM} & 0.683 & 0.487 & 0.206 & 0.017 & 0.316 & 0.468 & 0.598 \\ \hline
\multirow{5}{*}{\textbf{SDSS}} & \textbf{Meta*} & 0.866 & 0.813 & 0.704 & 0.459 & 0.804 & 0.813 & 0.838 \\ \cline{2-9}
 & \textbf{Meta} & 0.812 & 0.744 & 0.605 & 0.380 & 0.758 & 0.771 & 0.782 \\ \cline{2-9}
 & \textbf{Basic} & 0.761 & 0.680 & 0.547 & 0.339 & 0.696 & 0.697 & 0.717 \\ \cline{2-9}
 & \textbf{SVM$^r$} & 0.758 & 0.645 & 0.373 & 0.156 & 0.573 & 0.628 & 0.692 \\ \cline{2-9}
 & \textbf{SVM} & 0.747 & 0.556 & 0.164 & 0.023 & 0.319 & 0.502 & 0.647 \\ \hline

\end{tabular}
\end{table}

\begin{figure*}[t]
\vspace{-5pt}
\centering
    \subfigure[Results on $2$D]
    {\label{subf:2d} \includegraphics[width=0.49\columnwidth]{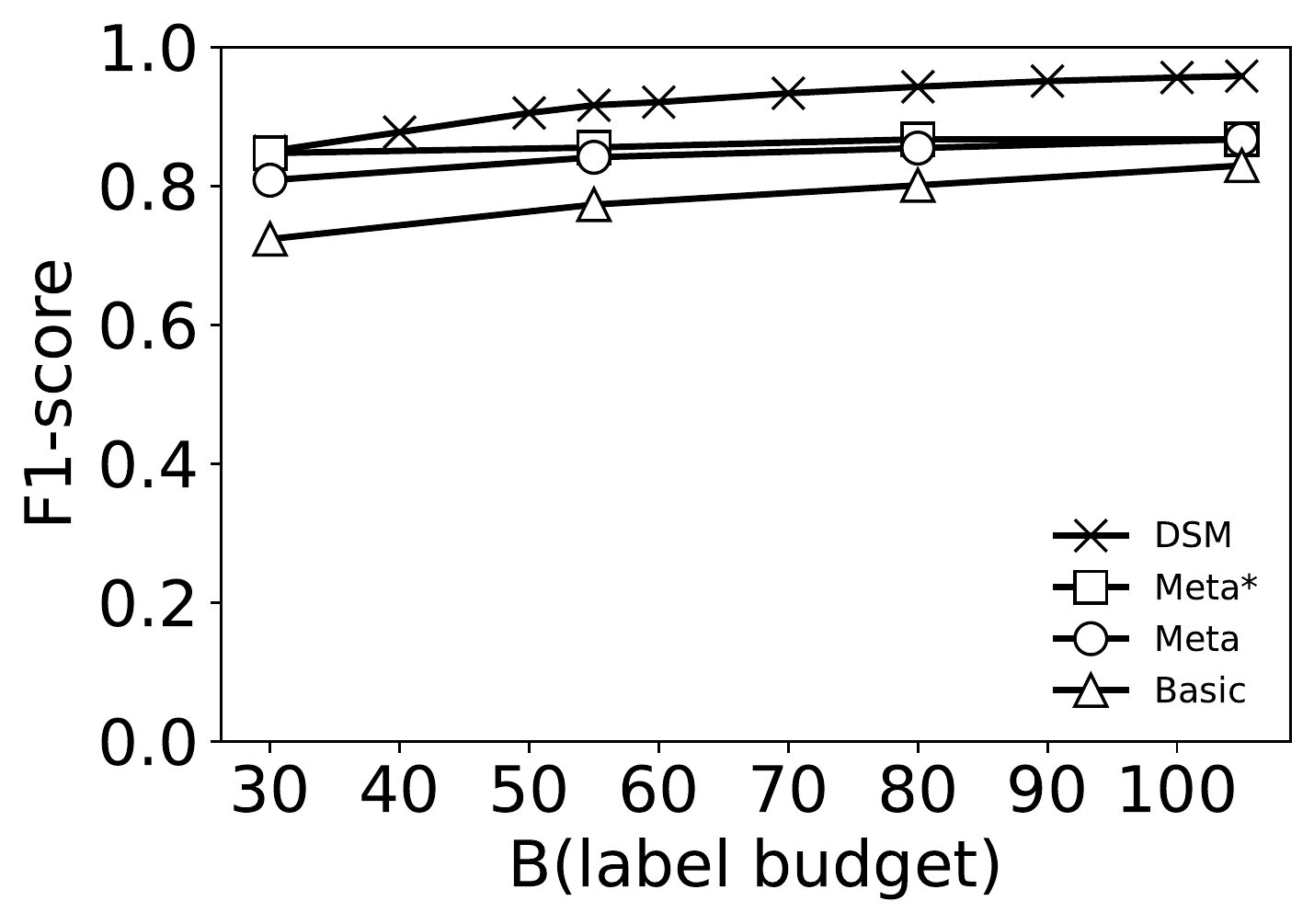}}
    \subfigure[Results on $4$D] {\label{subf:4d}\includegraphics[width=0.49\columnwidth]{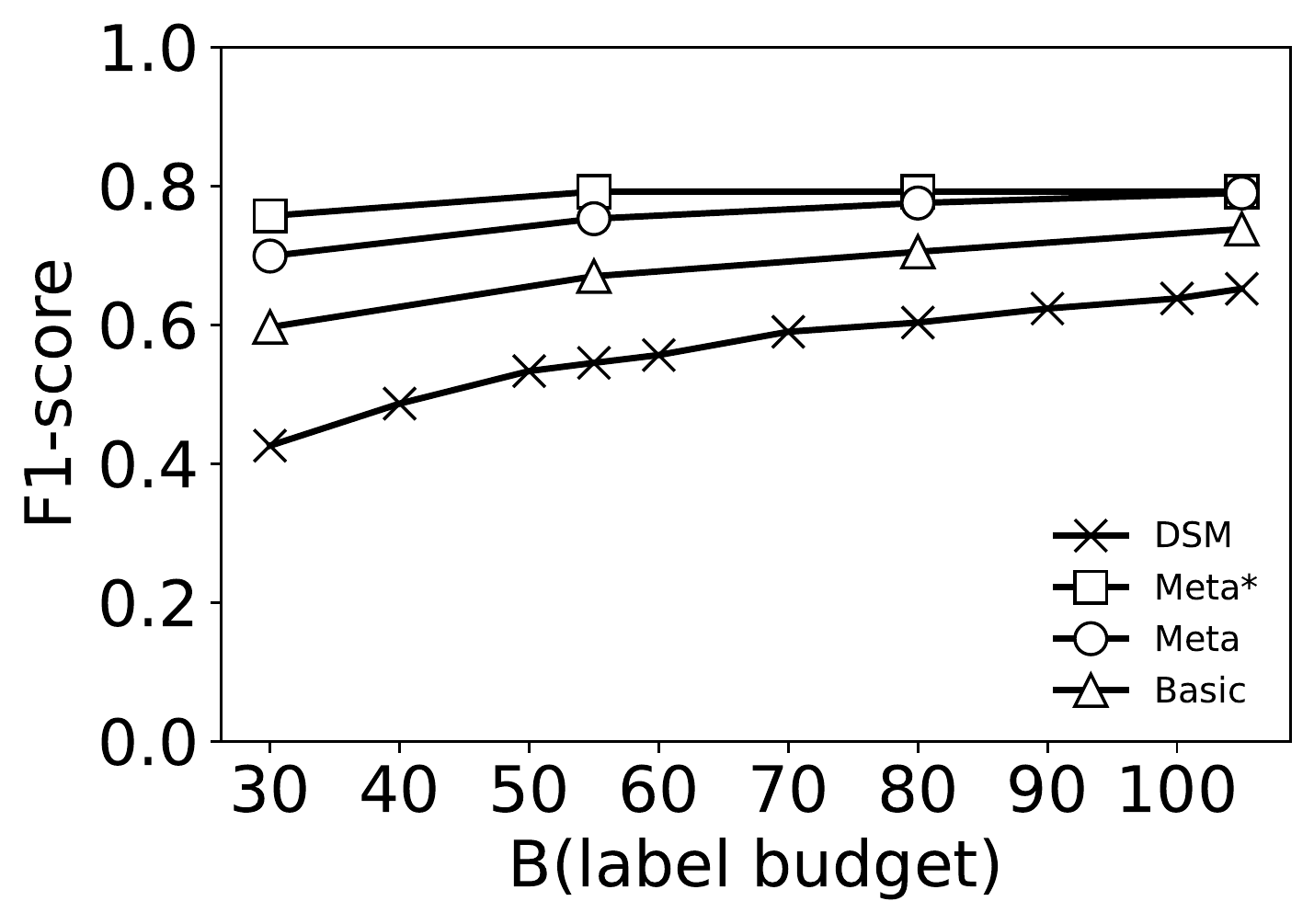}}
    \subfigure[Results on $6$D] {\label{subf:6d}\includegraphics[width=0.49\columnwidth]{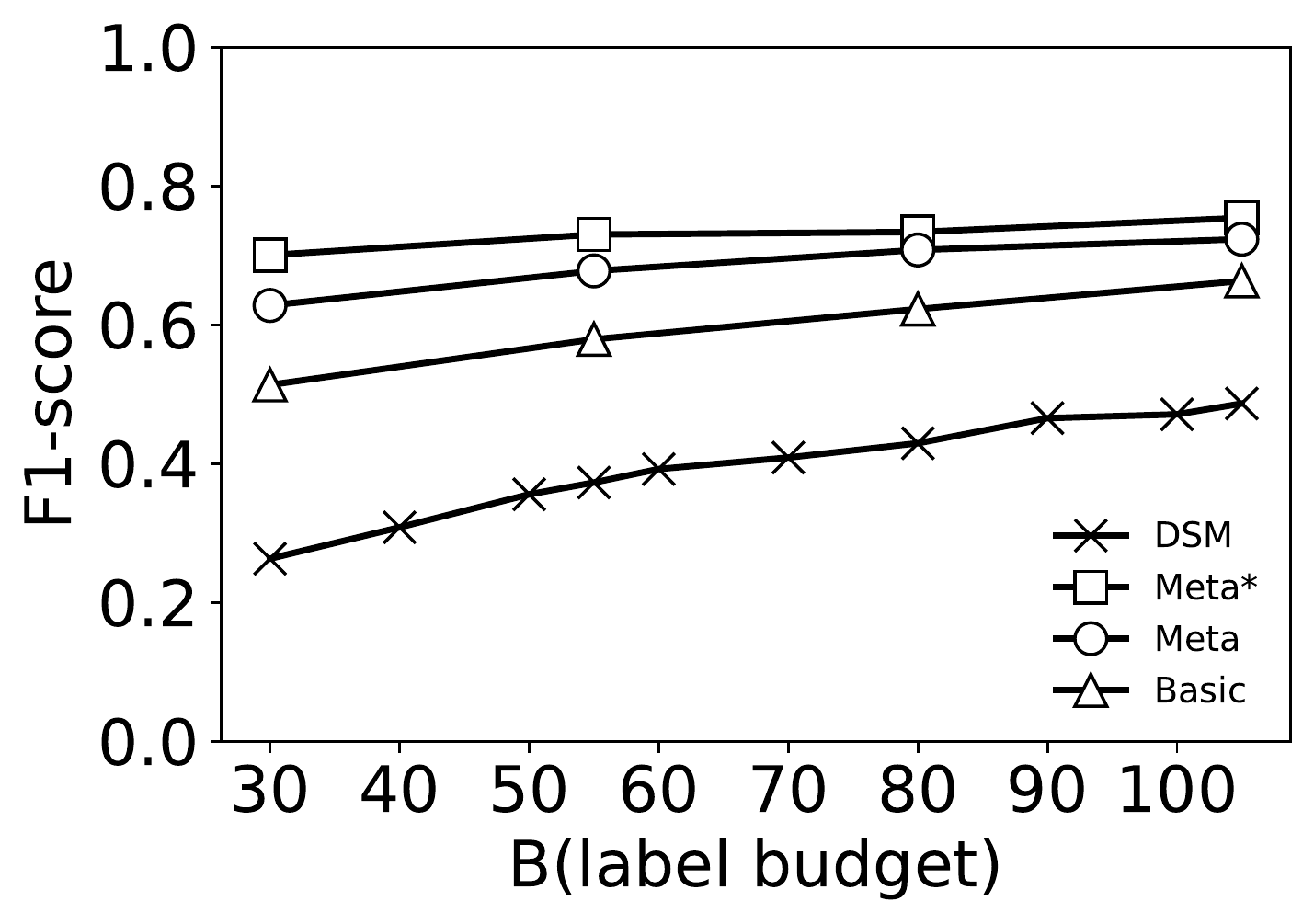}}
    \subfigure[Results on $8$D] {\label{subf:8d}\includegraphics[width=0.49\columnwidth]{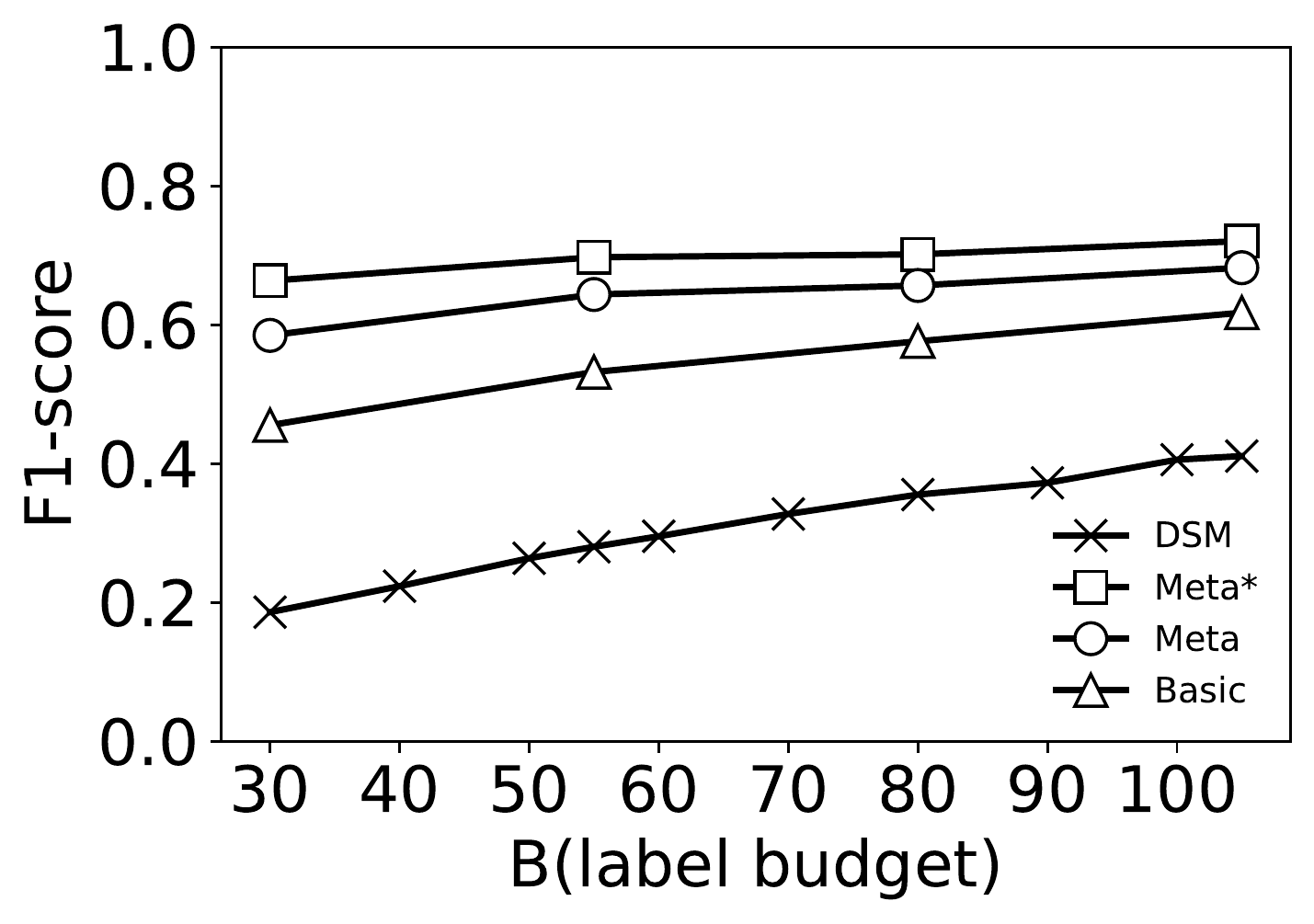}}
\label{fig:accvscost}
\vspace{-5pt}
\caption{Accuracy w.r.t. $B$ (SDSS, 4-8D)}
\vspace{-5pt}
\end{figure*}
\begin{figure*}[h]
\hspace{9pt}
\begin{minipage}[]{0.5\columnwidth}
\includegraphics[width=0.91\columnwidth]{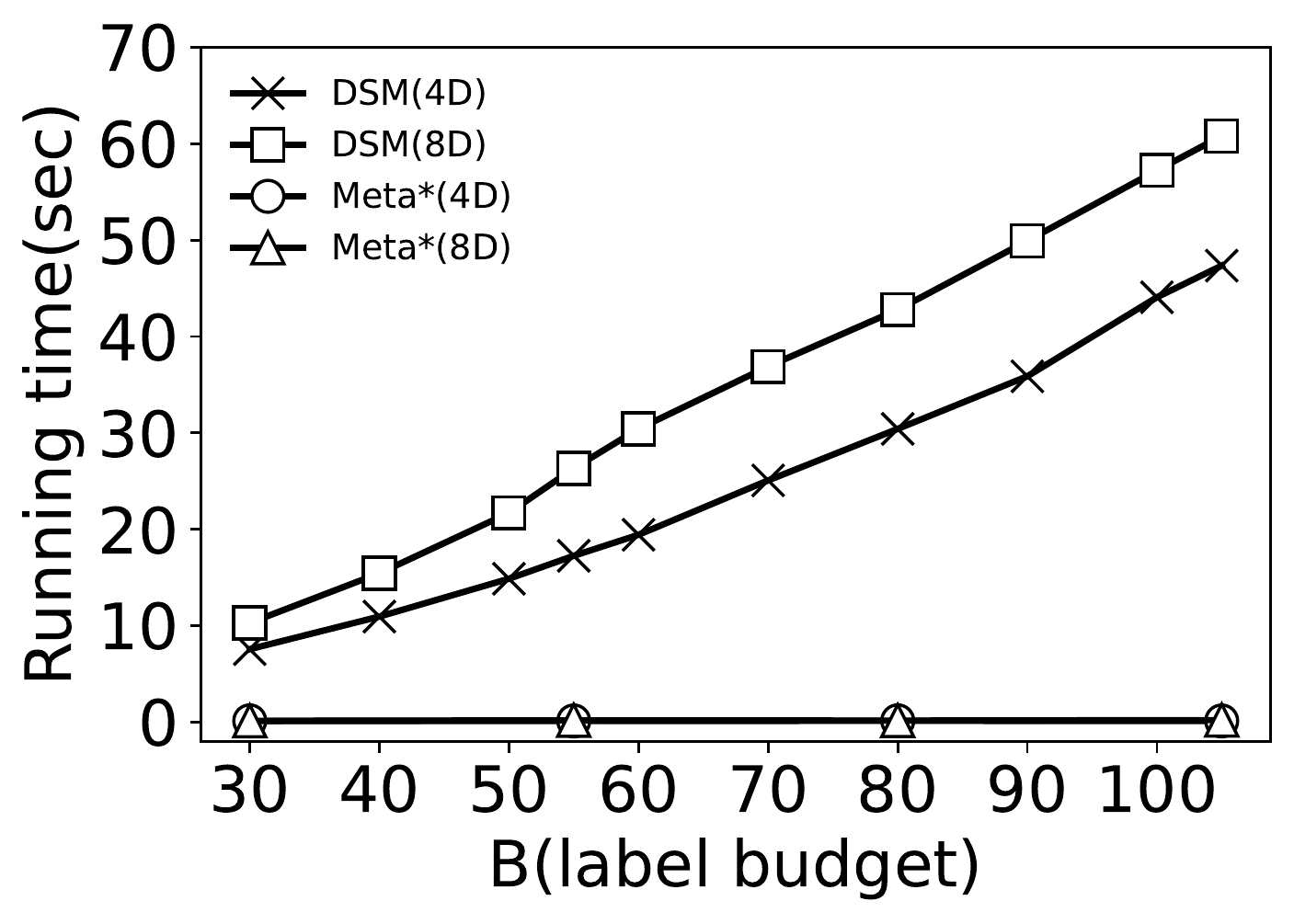}
\vspace{1pt}
\caption{Efficiency in Online Exploration}
\vspace{-5pt}
\label{subf:traineff}
\end{minipage}
\hspace{-15pt}
\begin{minipage}[]{1.5\columnwidth}
\subfigure[Accuracy w.r.t. $B$ (CAR)]{\label{subf:car}
\includegraphics[width = 0.31\columnwidth]{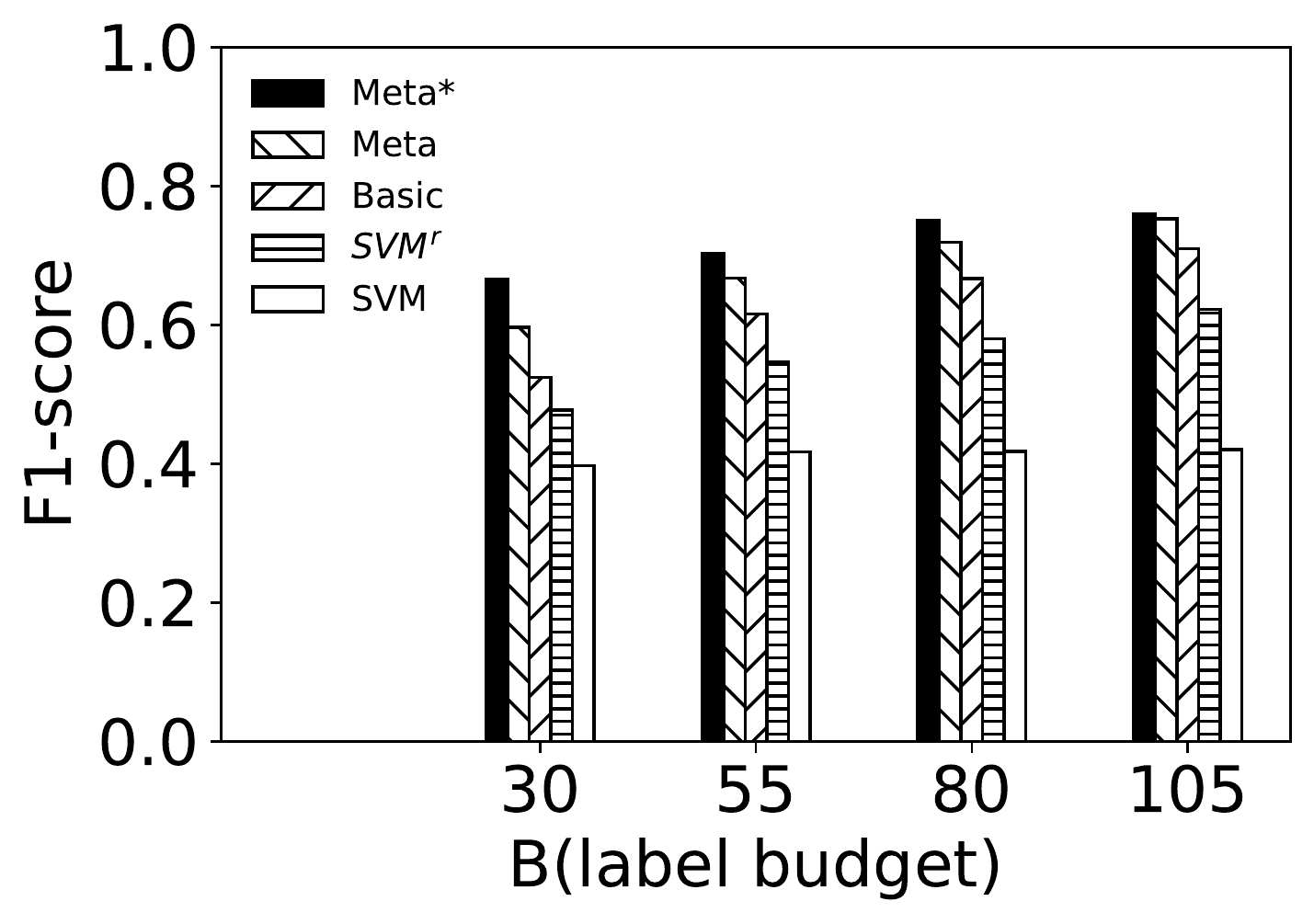}
}
\subfigure[Accuracy w.r.t. $B$ (SDSS)] {\label{subf:sdss}
\includegraphics[width = 0.31\columnwidth]{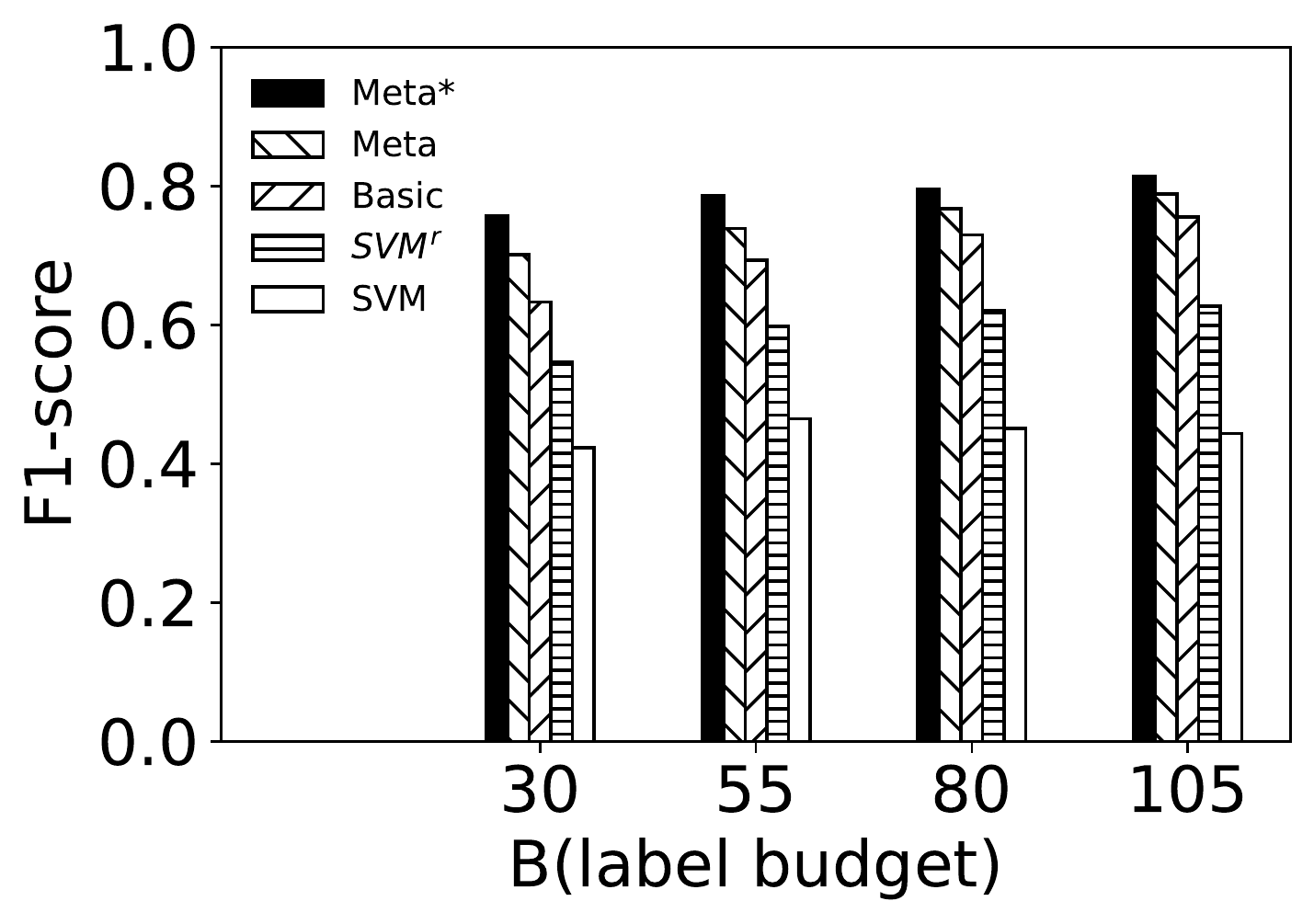}
}
\subfigure[Accuracy w.r.t. UIR Dim (SDSS)] {\label{subf:uis}
\includegraphics[width = 0.31\columnwidth]{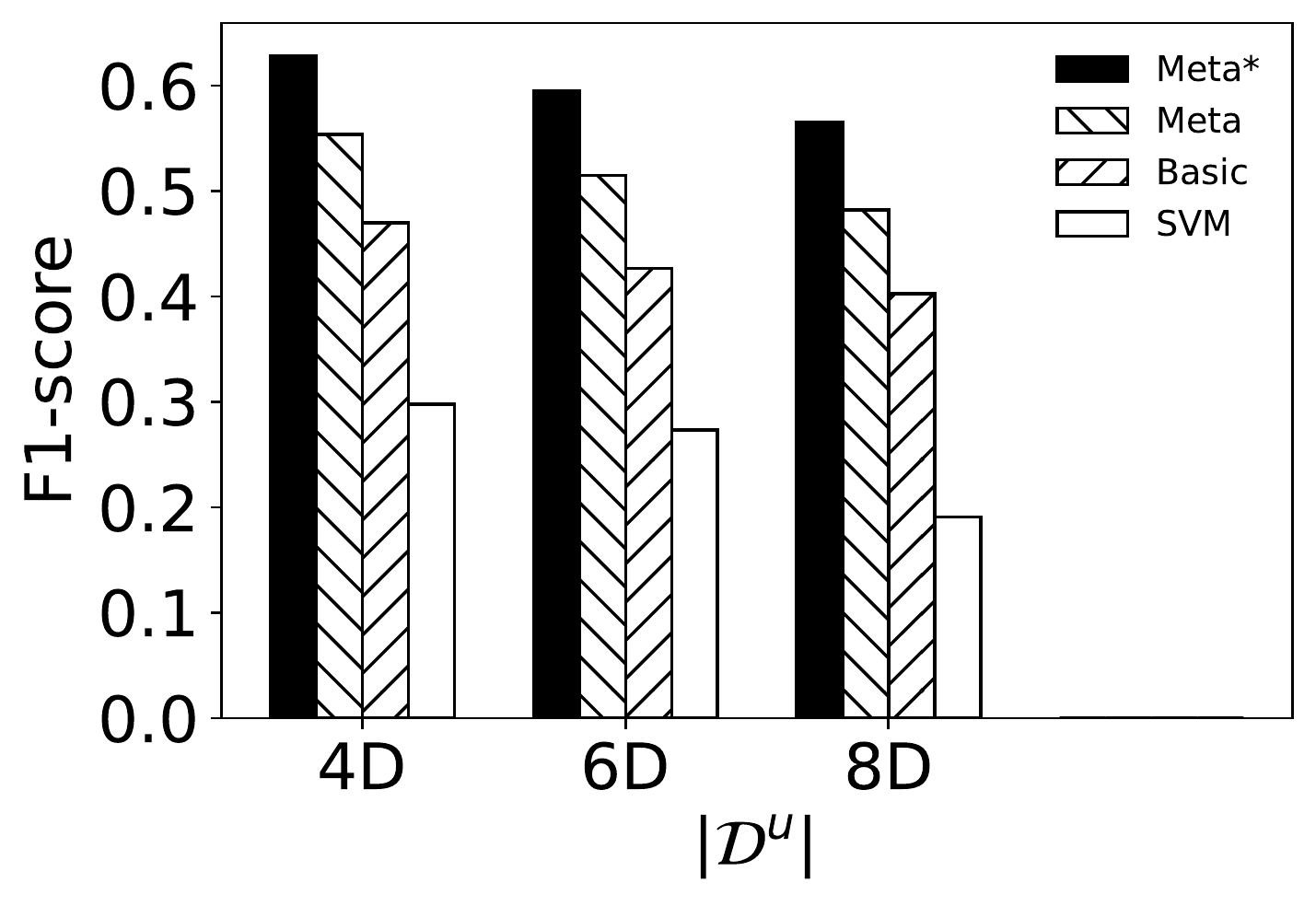}
}
\vspace{-15pt}
\caption{Performance on Generalized UIRs}
\vspace{-5pt}
\end{minipage}

\end{figure*}

\begin{figure*}[ht]
\vspace{-5pt}
\centering
\subfigure[GMM vs. JKC] {\label{subf:GMM_JKC}
\includegraphics[width=0.48\columnwidth]
{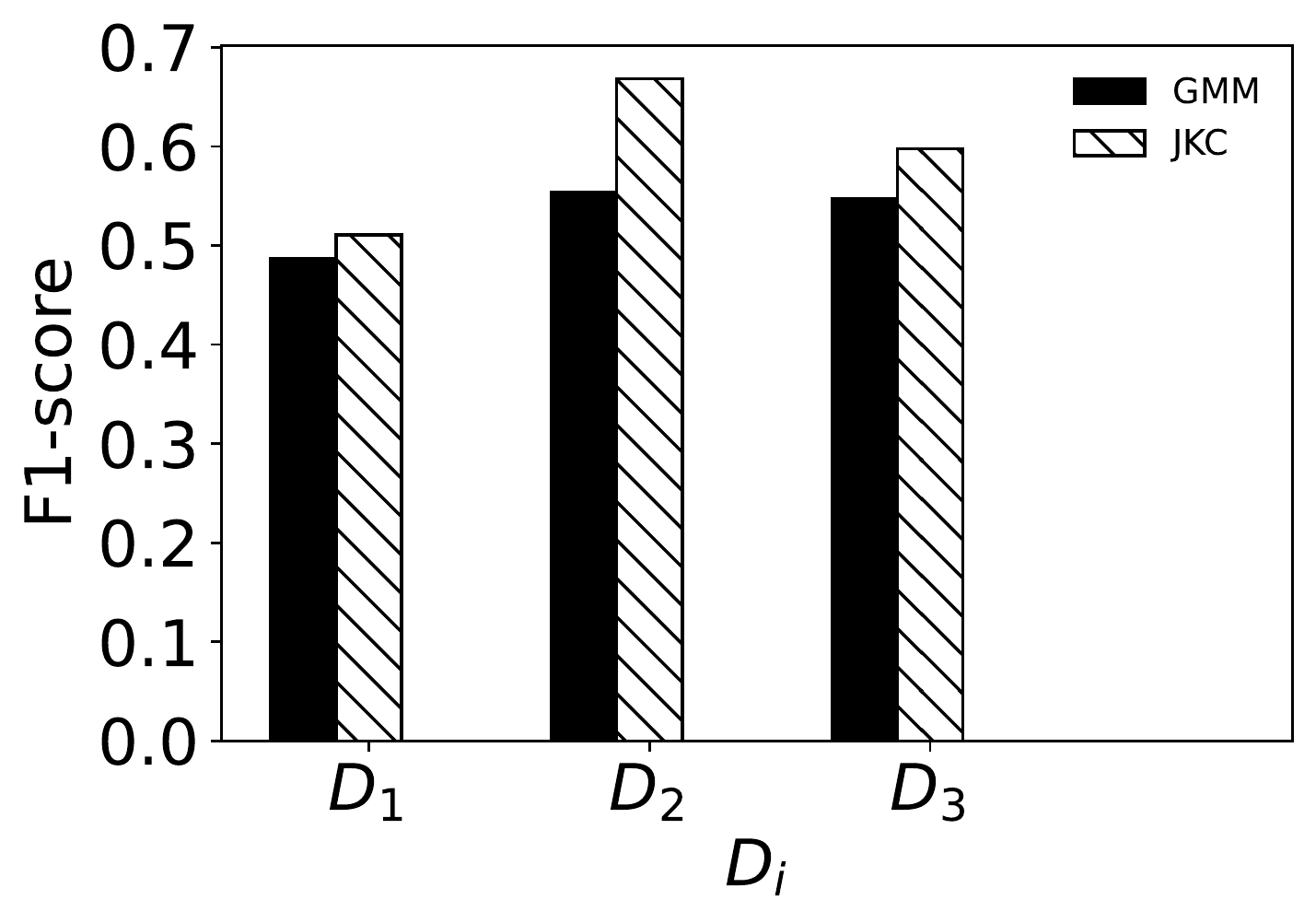}}
\subfigure[Pretraining Cost w.r.t. $|\mathcal{T}^M|$] {\label{subf:tasktime}\includegraphics[width=0.51\columnwidth]{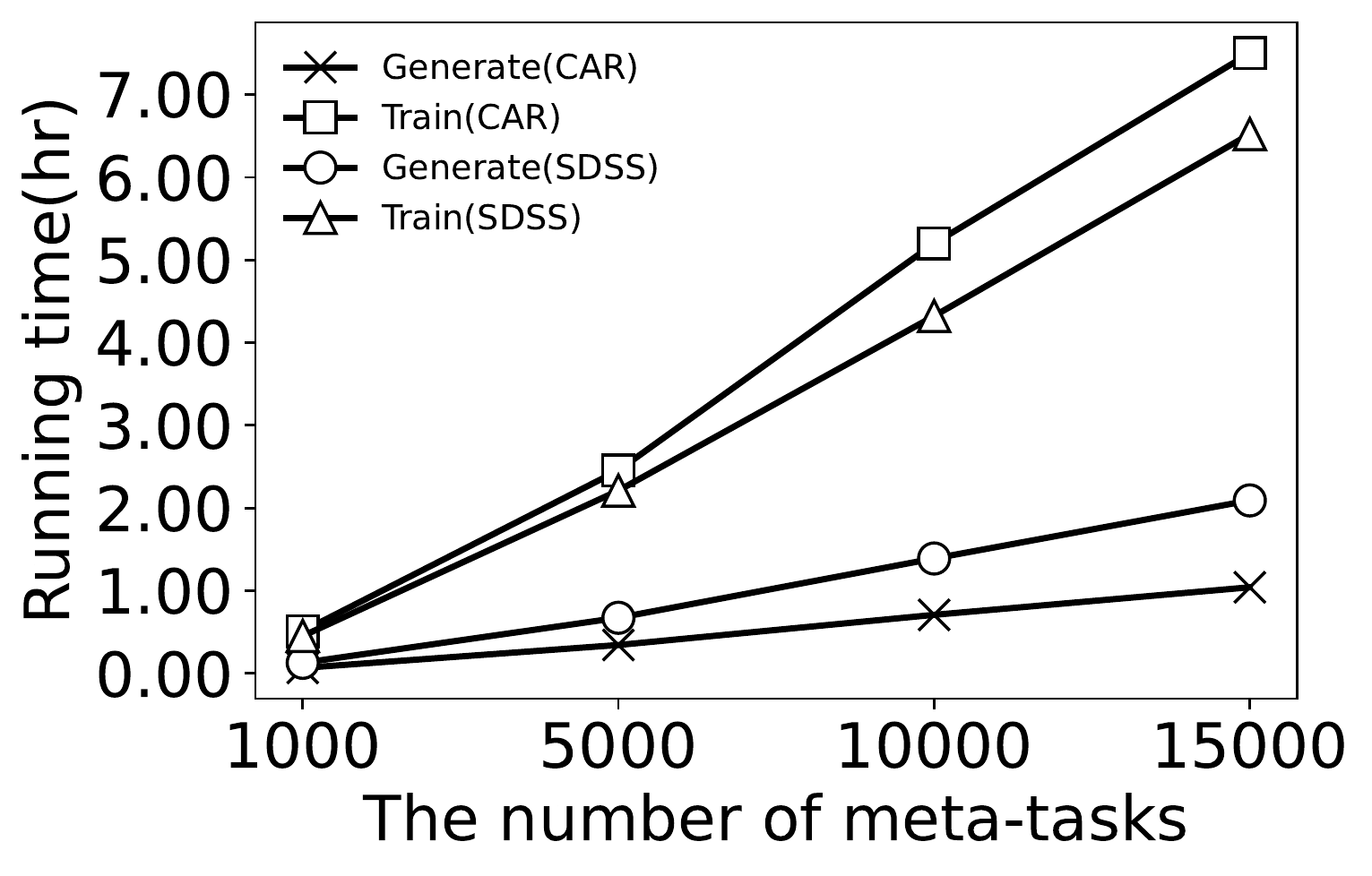}}
\subfigure[Accuracy w.r.t. $|\mathcal{T}^M|$] {\label{subf:taskacc}
\includegraphics[width=0.51\columnwidth]{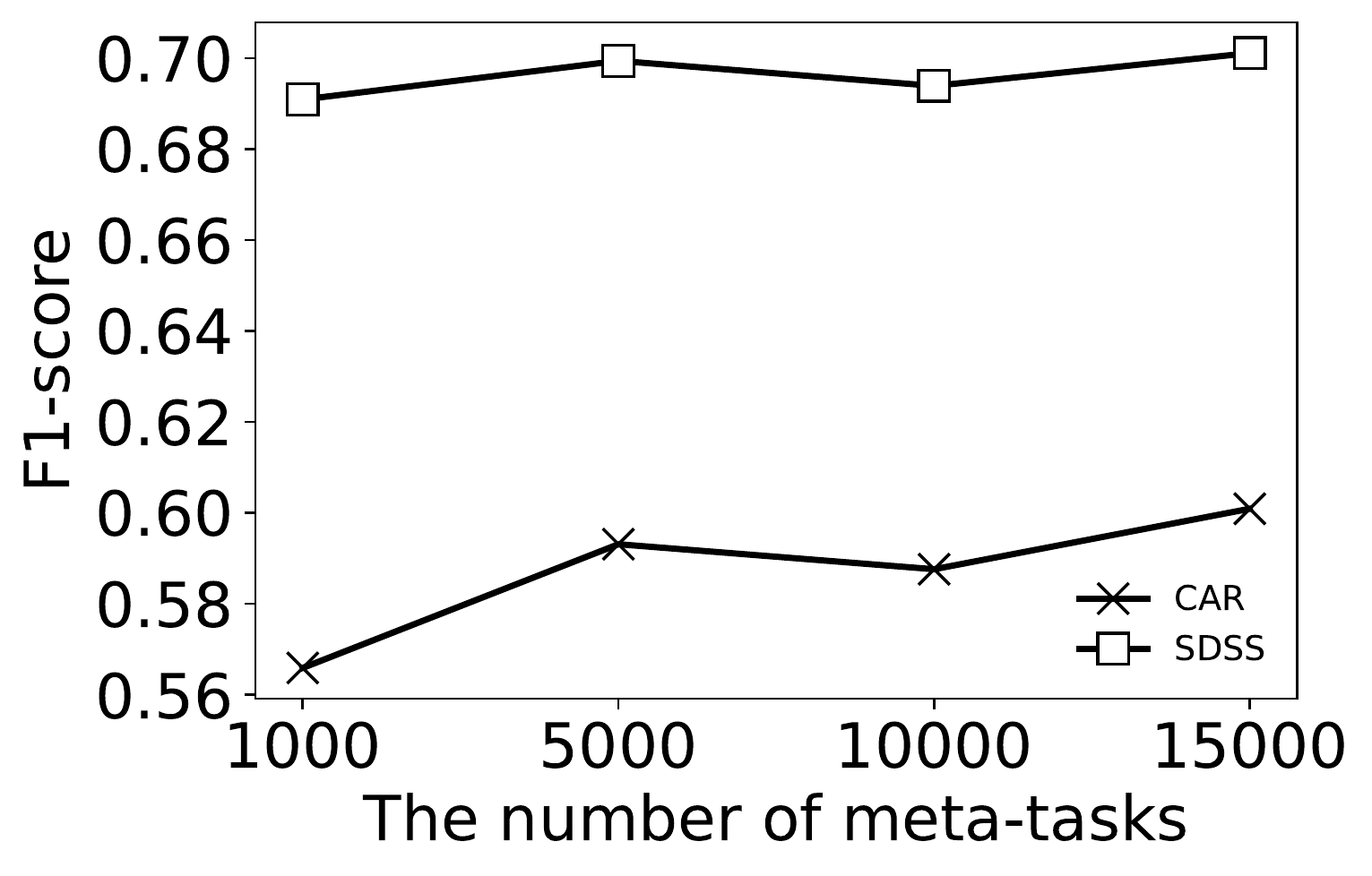}}
\subfigure[Accuracy w.r.t Learning Rate (Online exploration)] {\label{subf:lracc}\includegraphics[width=0.48\columnwidth]{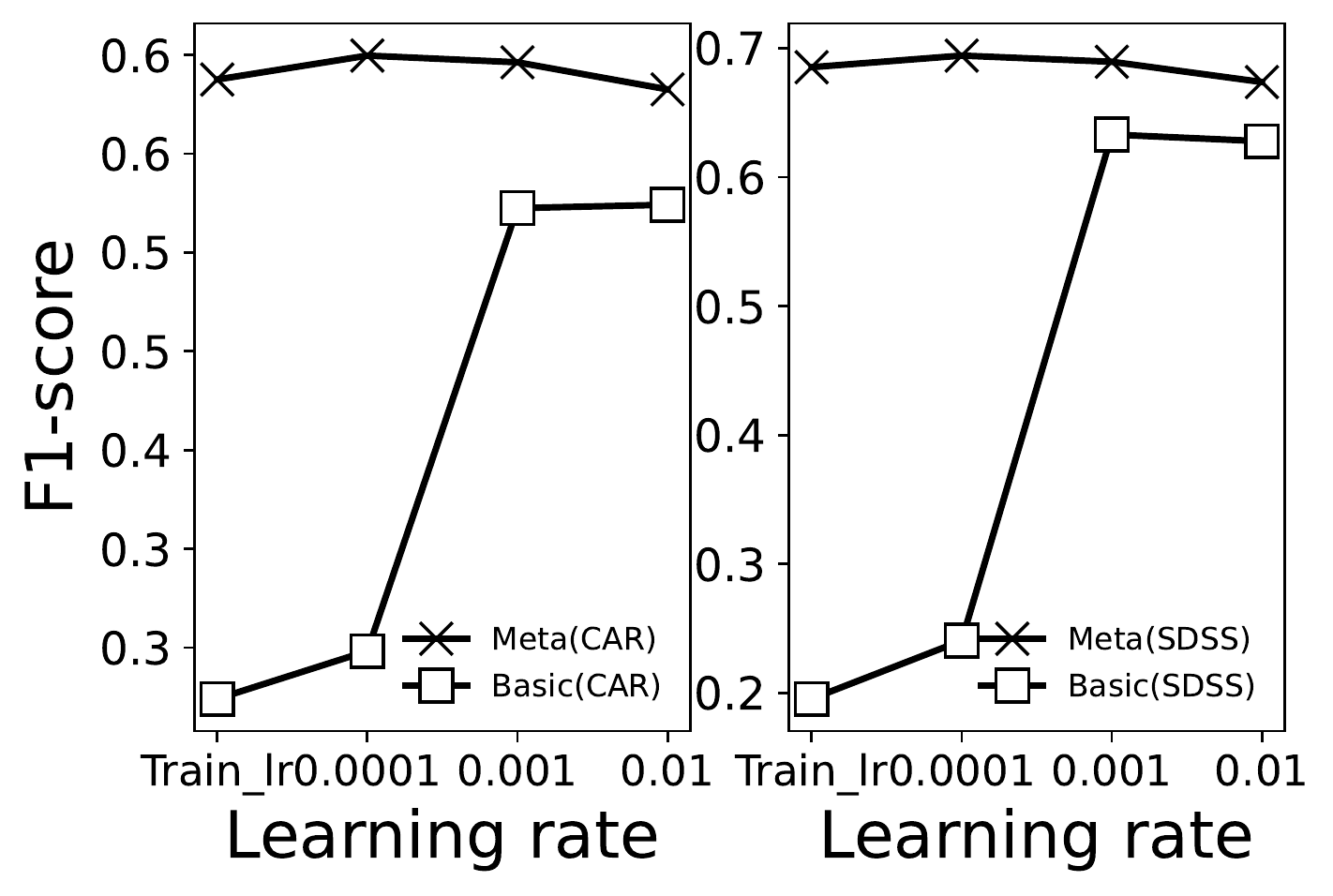}}
\vspace{-12pt}
\caption{Analysis}
\label{fig:analysis}
\vspace{-8pt}
\end{figure*}

\begin{table}[ht]
\vspace{-5pt}
\scriptsize
\caption{Modes of test benchmarks}
\vspace{-5pt}
\label{tab:test_typ}
\setlength{\tabcolsep}{2.5mm}
\centering
\begin{tabular}{cccccccc}
\hline
\textbf{Mode} & \textbf{M1} & \textbf{M2} & \textbf{M3} & \textbf{M4} & \textbf{M5} & \textbf{M6} & \textbf{M7} \\ \hline
\textbf{$\alpha$} & 4 & 4 & 4 & 4 & 1 & 2 & 3 \\ \hline
\textbf{$\psi$} & 20 & 15 & 10 & 5 & 20 & 20 & 20 \\ \hline
\end{tabular}
\vspace{-5pt}
\end{table}

We test the effect of exploration budgets $B$ over the accuracy in Figure 5 (a-d).
It shows that all methods' accuracy increases, if the given budget increases.
DSM performs better for 2D space, because experiments are done following the convexity and conjunctivity assumptions, which best fit its polytope-based optimization~\cite{10.14778/3275536.3275542}.
However, the performance of DSM degrades fast as the increase of dimensionality, which is consistent with the observation in Figure~\ref{fig:baseline}. For example, in $8$-dimensional space, the F1-score of DSM is below $20\%$, when $B=30$, and is below $41\%$ when $B=105$.
Compared to that, our methods, Meta and Meta*, scale well w.r.t. the number of dimensions.
The accuracy of Meta and Meta* dominates that of DSM. For example, when $|\mathcal{D}^u|=8$ and $B=30$, the F1-score of Meta* is $267\%$ of that of DSM.

We test the efficiency by collecting the runtime cost during the online exploration phase, for all competitors, in Figure~\ref{subf:traineff}.
It shows that the online training time of DSM increases almost linearly w.r.t. the given budget.
Also, if the number of dimensions increases, the training time takes longer.
For example, DSM takes about $50$ and $60$ seconds when $B$ equals $105$, on $4$- and $8$-dimensional spaces, respectively.
Compared to that, Meta*'s online exploration cost is two orders of magnitude lower because we save much cost by avoiding the online active learning process.
When the number of dimensions increases from $4$ to $8$, the online exploration cost only increases from $0.127$s to $0.130$s. It implies that our method has more potentials to provide the data exploration as a service, for a large number of users to access simultaneously.

\subsection{Performance on Generalized UIRs}
\label{subsec:ger_exp}

Our proposal supports UIRs, generalized from convex UIS to concave or even disconnected UIS in subspaces. 
We compare our proposals with SVM classifiers, since DSM degenerates into SVM classifiers, if UIS is not convex~\cite{10.14778/3275536.3275542}.
We consider a variant \textit{SVM$^r$} referring to using tabular data preprocessing in addition to SVM. All competitors are fed with the same set of initial training tuples for fair comparison.

The performance is tested on different UIS modes, which are randomly generated following the way of meta-task generation, as specified by two hyper-parameters, $\alpha$ and $\psi$.
To test the results on the combination of the two parameters, we first fix $\alpha$ to $4$ and vary $\psi \in\{20,15,10,5\}$, then fix $\psi$ to $20$ and vary $\alpha \in \{1,2,3\}$, so that we get $7$ UIS modes (M1-M7), as shown in Table~\ref{tab:test_typ}.
For each mode, we generate $100$ UISs for each subspace.
According to the statistics of generated
UISs, close to half of the UISs are concave or consist of separated regions.
Note that the meta-tasks used to train the meta-learners is only generated under $\alpha=4$, and $\psi=20$, and
all subsequent experiments follow this setting.

 Table~\ref{tab:acc_typ} shows the performance of each mode under labelling budget $B=30$. It shows that NN-based methods outperform SVM-based variants. Meta-learning based methods, Meta and Meta*, further improve basic in all testing modes.
 For example, for M4, the F1-score of Meta* is 164\% of that of SVM on CAR, showing the superiority of our method.
 Also, SVM$^r$ is better than SVM, due to the effectiveness of tabular data preprocessing.
 We then test the effect of meta-learning by comparing Meta and Basic. It shows that, from M5-M7 (CAR), the improvement of Meta over Basic is about $31$\%, $18$\%, and $12$\%, when $\alpha$ is set to $1$, $2$, and $3$.
 We also find that the improvement of Meta and Meta* over Basic is more significant, when $\alpha$ is small. Intuitively, a smaller $\alpha$ corresponds to a simpler task, thus the predicability can be higher. Compared to that, the trend over $\psi$ is relatively stable. Similar results are observed on SDSS. The above results show that the meta-learners trained under larger $\alpha$ and $\psi$ also perform well   on UIS configured by small $\alpha$ and $\psi$, so that we recommend larger valued $\alpha$ and $\psi$ for meta-task generation.

We also test the performance by varying  budget $B$ from $30$ to $100$ in
Figures~\ref{subf:car} and~\ref{subf:sdss}, for CAR and SDSS, respectively.  It shows that if the given budget increases, the accuracies of all methods except SVM increase. The reason is that when SVM handles a complex UIS, it is difficult to determine the appropriate hyper-parameters and kernel functions. In addition, our methods, Meta* and Meta, better predict complex UIS under a small $B$ with meta-knowledge. For example, on CAR, Meta with $B=55$ achieves the same performance as Basic with $B=80$.

Then, we show the performance w.r.t. dimensions of UIR, being generated by combining UISs from low-dimensional subspaces. UISs are generated according to Table~\ref{tab:test_typ}. Figure~\ref{subf:uis} examines the effect of accuracy over dimensions, with $B=30$. It shows that our method achieves relatively stable performance in different dimensions when UIR is complex.

\subsection{Analysis}
\label{subsec:anl}

\textit{\textbf{GMM vs. JKC.}} We study the effectiveness of tabular representations, JKC and GMM, as multi-mode feature models, in Figure~\ref{subf:GMM_JKC}.
If with GMM, the F1-score can be as high as $0.55$ for $2$D case. Basic integrates JKC and GMM representations, whose performance can be further improved (e.g, F1-score$=0.67$ for $2$D case).
Without JKC and GMM, the model can hardly be trained and used, with a F1-score even much lower than baselines.

\textbf{\textit{Pre-training Cost.}} We investigate the performance of runtime efficiency and accuracy w.r.t. the number of meta-tasks in Figures~\ref{subf:tasktime} and~\ref{subf:taskacc}.
The runtime cost refers to two parts, the generation time for meta-tasks, and training time. Both of the two parts are linearly proportional to the number of meta-tasks, as shown in Figure~\ref{subf:tasktime}.
Meanwhile, we find that the runtime cost does not depend on the dataset size. For example, CAR takes only half of the data size of SDSS, but the training time is only 12\% less.
We also test the accuracy w.r.t. the number of meta-tasks in Figure~\ref{subf:taskacc}.
It shows that for both datasets, the accuracy is not sensitive to the number of meta-tasks, except the number of tasks is low, e.g., $|\mathcal{T}^M|=1,000$.
There exist some fluctuations in Figure~\ref{subf:taskacc}, which are consistent with the consensus \cite{hospedales2020metalearning} that the meta-learning performance w.r.t. the number of meta-tasks follows a gradual transition from positive correlation to fluctuation with stationarity.
So, we can do an early stop for meta-training by finding a ``sweet point'' of accuracy and efficiency. According to Figures~\ref{subf:tasktime} and~\ref{subf:taskacc}, when $|\mathcal{T}^M|=5,000$, the accuracy is almost at the peak, while the training runtime cost is low.

\textbf{\textit{The Effect of Meta-Learning.}} We study the effect of meta-learning by comparing Basic and Meta, under different values of learning rates.
We only compare the two to show the effectiveness of meta-learning, by eliminating the influence of other factors.
The learning rate refers to the step size at each iteration, while moving towards the minimum of a loss function during optimization.
For offline training, a small learning rate $(0.00005)$ is chosen to conservatively and deliberately capture the meta-knowledge.
For online exploration, a large learning rate is preferred for fast converging to UIR.
The result is shown in Figure~\ref{subf:lracc}, where Meta steadily outperforms Basic.
The improvement is achieved because Meta is equipped with meta-knowledge, in form of good initial parameters, so the sensitiveness to learning rates is low and the performance is stable. So, Meta can achieve good results at a small learning rate.
For example, when learning rate is $0.0001$,
the F1-score of Meta is $0.7$, but the F1-score of Basic is only $0.25$, which is 64\% lower, on SDSS. Similar results are observed on CAR.

\section{Conclusion}
\label{sec:con}

In this paper, we study the problem of interactive data exploration by proposing a ``learn-to-explore'' framework. The framework leverages meta-learning based neural network classifiers, which are pre-trained by automatically generated meta-tasks in an unsupervised manner, and are fast adapted to optimal parameters during the online exploration.
The framework can be plugged to existing IDE systems by providing good initial parameters for classifiers, thus yielding good accuracy and efficiency.
To implement such a framework, we study a set of techniques, including meta-task generation, meta-training, etc. Experiments on real datasets show that our proposal outperforms existing solutions in terms of accuracy and efficiency.

\newpage

\bibliographystyle{IEEEtran}
\bibliography{ICDE.bib}

\end{document}